\documentclass[sigconf]{acmart}

\usepackage{subfigure}
\AtBeginDocument{%
  \providecommand\BibTeX{{%
    \normalfont B\kern-0.5em{\scshape i\kern-0.25em b}\kern-0.8em\TeX}}}

\setcopyright{acmcopyright}
\copyrightyear{2021}
\acmYear{2021}
\acmDOI{10.1145/1122445.1122456}
\usepackage{multirow}
\acmConference[WSDM 2021]{Woodstock '18: ACM Symposium on Neural
  Gaze Detection}{March 08--12, 2021}{Jerusalem, Israel}
\acmBooktitle{Woodstock '21: ACM Symposium on Neural Gaze Detection,
  March 08--12, 2021, Jerusalem, Israel}
\acmPrice{15.00}
\acmISBN{978-1-4503-XXXX-X/18/06}

\begin{document}

\title{Learning User Representations with Hypercuboids for Recommender Systems}
%Representing Users as Hypercuboids for Recommender Systems

%%
%% The "author" command and its associated commands are used to define
%% the authors and their affiliations.
%% Of note is the shared affiliation of the first two authors, and the
%% "authornote" and "authornotemark" commands
%% used to denote shared contribution to the research.
% \author{Shuai Zhang\star, \text{Huoyu Liu}\clubsuit, \text{Aston Zhang}\spadesuit, \text{Yue Hu}\clubsuit, \and \text{Ce Zhang}\star,  \text{Yumeng Li}\clubsuit, \text{Tanchao Zhu}\clubsuit, \text{Shaojian He}\clubsuit, \text{Wenwu Ou}\clubsuit}
% \author{Shuai Zhang, Huoyu Liu, Aston Zhang, Yue Hu,  Ce Zhang, Yumeng Li, Tanchao Zhu, Shaojian He, Wenwu Ou}

% \author{\spacedlowsmallcaps{John Smith* \& James Smith\textsuperscript{1}} 
% \spacedlowsmallcaps{John Smith* \& James Smith\textsuperscript{1}}}
% \orcid{1234-5678-9012}
% \author{Shuai Zhang}
% \affiliation{%
%   \institution{\star \text{Department of Computer Science, ETH Zurich} 
%   }
% %  \institution{ \\
% %   \clubsuit \text{University of Illinois at Urbana-Champaign} 
% %   }
% %   \city{Zurich}
% %   \country{Switherland}
% }
% \affiliation{%
%   \institution{\clubsuit \text{Alibaba Group} 
%   }
% }
% \affiliation{%
%   \institution{\spadesuit \text{University of Illinois at Urbana-Champaign} 
%   }
% }
% \email{shuai.zhang@inf.ethz.ch,  huoyu.lhy@alibaba-inc.com}

\author{Shuai Zhang}
\affiliation{%
  \institution{Department of Computer Science, ETH Zurich}
 }
\email{shuai.zhang@inf.ethz.ch}

\author{Huoyu Liu}
\affiliation{%
  \institution{Alibaba Group}
 }
\email{huoyu.lhy@alibaba-inc.com}

\author{Aston Zhang}
\affiliation{%
  \institution{University of Illinois at Urbana-Champaign}
%   \city{Champaign}
%   \country{United States}
}

\author{Yue Hu}
\affiliation{%
  \institution{Alibaba Group}
 }
 
 \author{Ce Zhang}
\affiliation{%
  \institution{Department of Computer Science, ETH Zurich}
%   \city{Zurich}
%   \country{Switherland}
}

 \author{Yumeng Li, Tanchao Zhu, Shaojian He, Wenwu Ou }
\affiliation{%
  \institution{Alibaba Group}
%   \city{Hang Zhou}
%   \country{China}
  }

% \email{cpalmer@prl.com}

% \author{John Smith}
% \affiliation{\institution{The Th{\o}rv{\"a}ld Group}}
% \email{jsmith@affiliation.org}

% \author{Julius P. Kumquat}
% \affiliation{\institution{The Kumquat Consortium}}
% \email{jpkumquat@consortium.net}

%%
%% By default, the full list of authors will be used in the page
%% headers. Often, this list is too long, and will overlap
%% other information printed in the page headers. This command allows
%% the author to define a more concise list
%% of authors' names for this purpose.
\renewcommand{\shortauthors}{}

%%
%% The abstract is a short summary of the work to be presented in the
%% article.
\begin{abstract}
Modeling user interests is crucial in real-world recommender systems. In this paper, we present a new user interest representation model for personalized recommendation. Specifically, the key novelty behind our model is that it explicitly models user interests as a hypercuboid instead of a point in the space. In our approach, the recommendation score is learned by calculating a compositional distance between the user hypercuboid and the item. This helps to alleviate the potential geometric inflexibility of existing collaborative filtering approaches, enabling a greater extent of modeling capability. Furthermore, we present two variants of hypercuboids to enhance the capability in capturing the diversities of user interests. A neural architecture is also proposed to facilitate user hypercuboid learning by capturing the activity sequences (e.g., buy and rate) of users. We demonstrate the effectiveness of our proposed model via extensive experiments on both public and commercial datasets. Empirical results show that our approach achieves very promising results, outperforming existing state-of-the-art.

\end{abstract}

\ccsdesc[500]{Information systems~Recommender systems}
\ccsdesc[500]{Computing methodologies~Neural networks}

\keywords{Recommender Systems, Hypercuboids, User Representation}

%%
%% This command processes the author and affiliation and title
%% information and builds the first part of the formatted document.
\maketitle

\section{Introduction}

We propose a novel recommendation approach with hypercuboids. The hypercuboid (also known as hyperrectangle) is a generalization of a three-dimensional cuboid to four or more dimensions. More specifically, we use the expressive hypercuboids to represent users in order to capture their complex and diverse interests. Recommendations are made based on the relationships between user hypercuboids and items. 

This work is motivated by the recent popularity and growing importance of recommender systems in real-world applications (e.g., e-commerce  and online entertainment platforms). As an effective tool to ameliorate information overload and generate income, its importance cannot been overestimated. A major step in recommender systems is finding the right items that suit users' preferences and interests. However, obtaining informative user representations is a challenging task as users can have a wide range of interests and different tastes concerning different categories of items. The massive amount and diversity of item sets in real world applications make the case even worse. 

\begin{figure}[t]
\begin{center}
\subfigure{
\includegraphics[width=0.7\linewidth]{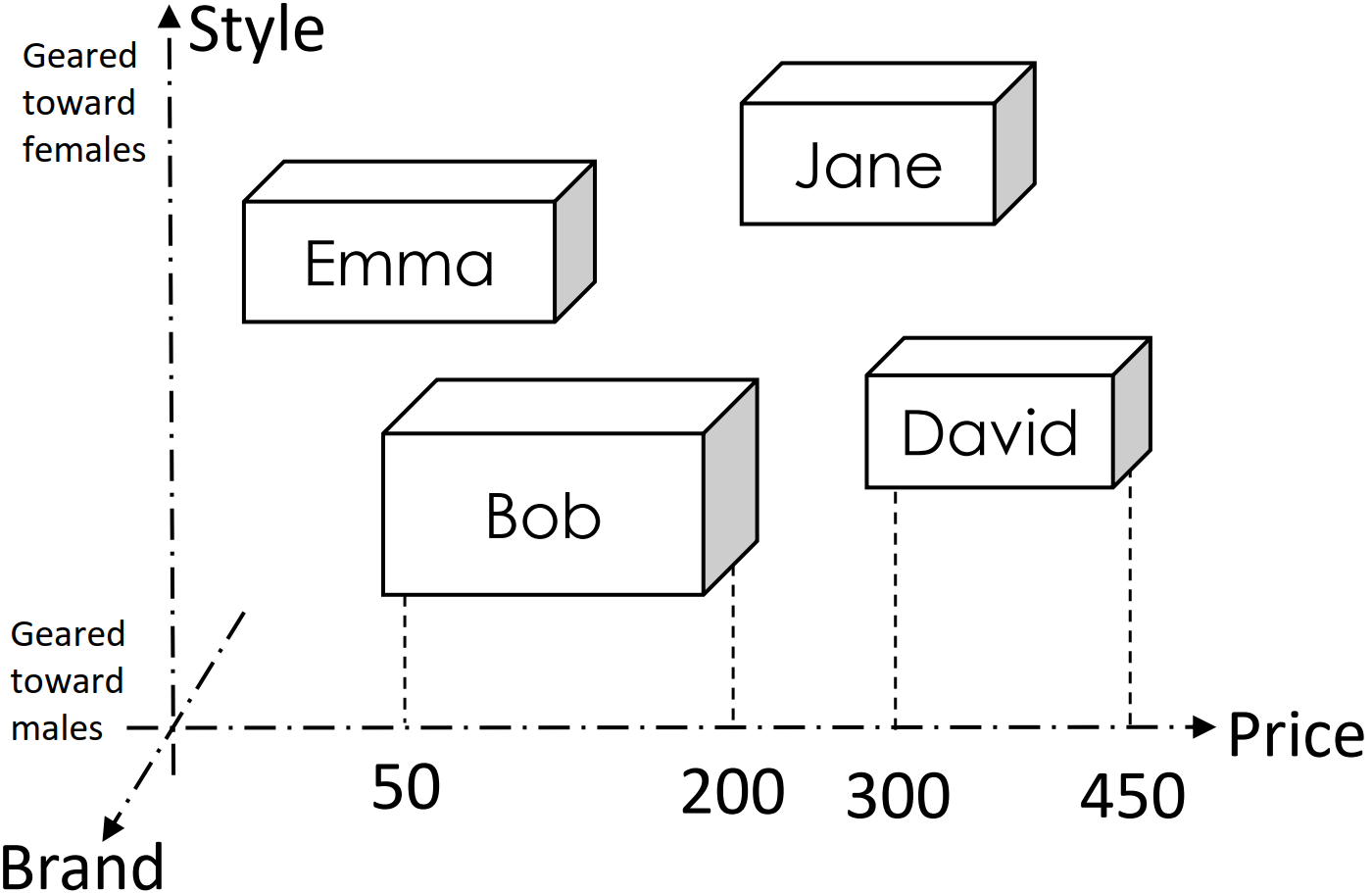}
\label{fig:example}}
\caption{Representing Users as Hypercuboids in 3-D space. Interest on each dimensional can span a scope.}
\label{example}
\end{center}
\vspace{-1em}
\end{figure}

Common user modeling approaches aim to represent users as points in vector space. For example, the standard factorization model~\cite{koren2009matrix} factorizes the large interaction matrix into user and item latent embeddings in  low-dimensional space. As a result, recommendation is reduced to identifying items which are near or similar to the user in that space. Despite its popularity and wide applications, this paradigm faces weaknesses. On one hand, embedding the disparate and diverse interests into single points is problematic~\cite{weston2013nonlinear}. Intuitively, users are likely to have different tastes and considerations for items from different categories or domains.  On the other hand, for most factors, users have preferred ranges spread. For example, consumers usually have their acceptable price range. It is inexpressive to model such range with a single value. Clearly, such systems are geometrically restrictive.  

Thus, we need a much richer latent representation system to overcome these limitations. To this end, we propose a new paradigm that represents users as hypercuboids.  We go beyond point embeddings, adopting the more expressive hypercuboids to effectively capture the intricate and complex interests of users. A hypercuboid is a subspace that encapsulates innumerable points, enabling more representation capacities than a single point. Moreover, the preference range of each factor (e.g., accepted price) can be naturally modelled with the edge of hypercuboids. 

Overall, the basic idea is that: for each user, we learn single or multiple hypercuboids to represent the user's interests. The relationship between the user hypercuboids and item points is captured by a re-designed distance measure, where both distances outside and inside the hypercuboid are considered. Ideally, the user hypercuboid will allow for a greater degree of flexibility and expressiveness.

%  Since the hypercuboid has the interior, it is natural to position items that the user likes inside the hypercuboid and items that the user dislikes outside the hypercuboid.

The main contributions of this work are as follows:

\begin{itemize}
    \item  We present a novel user representation method for personalized recommendation. For the first time, we adopt hypercuboids to represent users' complicated user preferences or interests. Two variants are proposed to further enhance the representation capacity in capturing the diversity of user interests.
    \item  We design a distance function to measure the distance between items and user hypercuboids. This measure provides improvements in terms of flexibility and model expressiveness of the algorithms.  
    \item We conduct experiments on six datasets. Our model achieves the best performances on all the benchmark datasets. We investigate the inner workings of our model to gain a better intuition pertaining to the model's high performance. We also provide case study on real-world datasets to show the effectiveness of the proposed approach. 
\end{itemize}

% We present LRML (Latent Relational Metric Learning), a
% novel, end-to-end neural network architecture for collaborative filtering and ranking on implicit interaction data. For
% the first time, we adopt user and item specific latent relation vectors to model the relationship between user-item
% interactions.
% • We propose a novel Latent Relational Attentive Memory (LRAM)
% module in order to generate the latent relation vectors. The
% LRAM module provides improvements in terms of flexibility
% and modeling capability of the algorithm. Moreover, the neural attention also gives greater insight and interpretability
% of the model.

\section{The Proposed Method}
In this section, we formulate the hypercuboid representation for user interest modeling, introduce two variants of hypercuboid, and present the overall architecture for hypercuboid learning.

Our work is concerned with the recommendation task with implicit feedback. In a standard setting, we suppose that there are a set of users, $u \in \{1, ..., |\mathcal{U}|\}$ and a set of items,  $i \in \{1, ..., |\mathcal{I}|\}$. The goal is to generate a set of recommendable items for a user given existing feedback, typically, implicit feedback such as clicks, buys, likes. We set the interaction between user $u$ and item $i$, $y_{ui}$ to $1$ if the implicit feedback exists, otherwise, $y_{ui}=0$, where $y_{ui}$ is an entry of the interaction matrix $\textbf{Y} \in \mathbf{R}^{|\mathcal{U}| \times |\mathcal{I}|}$. In addition, time steps will also be considered in the following subsections.

\subsection{Representing Users as Hypercuboids}
Hypercuboid is an analogue of a cuboid in four or more dimensions. For the convenience of computation, we define a hypercuboid with a center and an offset, both with the same dimensions.  Formally, we operate on $\mathbb{R}^d$. Let $\mathbf{c} \in \mathbb{R}^d$ represent the center of a hypercuboid and $\mathbf{f} \in \mathbb{R}^d_{0+}$ denote the non-negative offset. The offset is used to determine the lengths of the edges of the hypercuboid. A hypercuboid is defined as:
\begin{equation}
    Hypercuboid \equiv \{\mathbf{p} \in \mathbb{R}^d:  \mathbf{c} - \mathbf{f} \preceq \mathbf{p} \preceq \mathbf{c} + \mathbf{f} \},
\end{equation}
where $\mathbf{p}$ denotes the point which is inside the hypercuboid.

In our framework, every user is modeled with such a hypercuboid.  We use bold uppercase letters $\mathbf{C} \in \mathbb{R}^{|\mathcal{U}| \times d}$ and $\mathbf{F} \in \mathbb{R}^{ | \mathcal{U} | \times d}_{0+}$ to denote the centers and offsets respectively for all the users. As such, a user is represented by a hypercuboid $\text{user}_u(\mathbf{C}_u, \mathbf{F}_u)$. Items are points and are embedded by $\mathbf{V} \in \mathbb{R}^{ | \mathcal{I} |  \times d}$. Obviously, an item point can be outside, inside, or on the surface of the user hypercuboid. A proper method is needed to measure the relationships between users and items. Meanwhile, the characteristics of hypercuboid should be retained. 

% \begin{figure}[t]
% \centering
% \includegraphics[width=0.65\linewidth]{KDD2020/Hcrec.pdf}
% \centering\caption{Limitation of point embeddings, $item_1$ and $item_2$ are not distinguishable when user is a point. Representing the user as a hypercuboid can effectively address this problem. }
% \label{fig:hypercubioid}
% \end{figure}

\begin{figure}[t]
\begin{center}
\subfigure{
\includegraphics[width=0.8\linewidth]{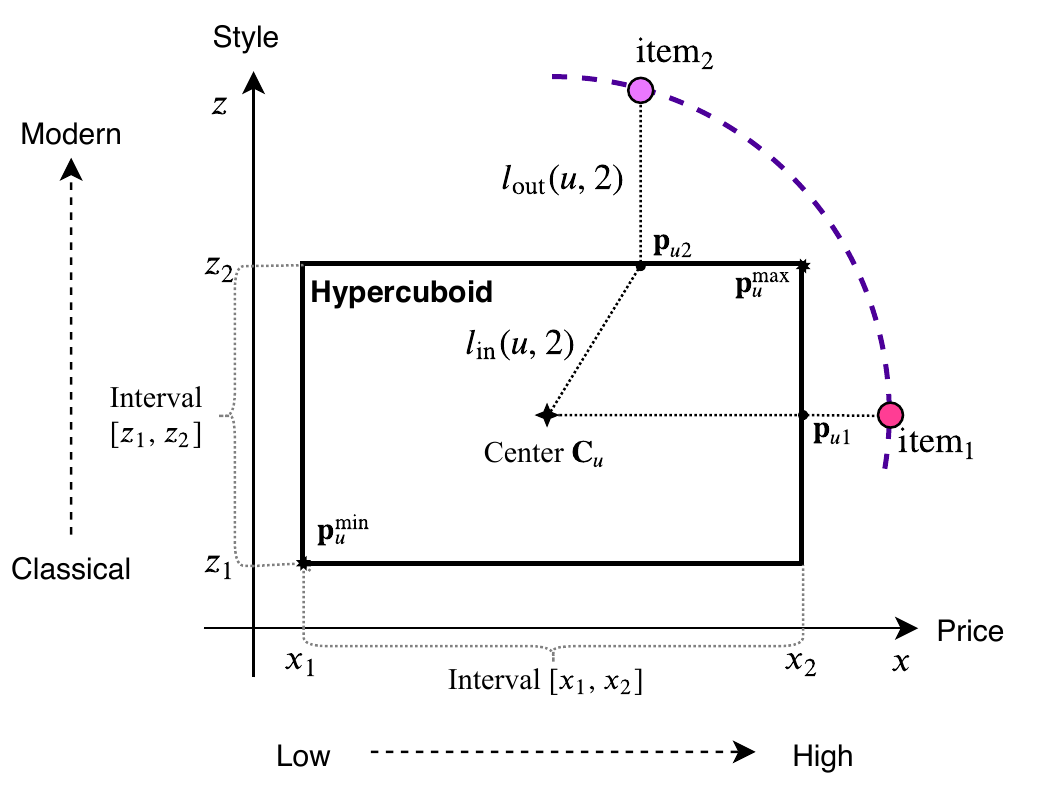}
}
\caption{Limitation of point embeddings, $item_1$ and $item_2$ are not distinguishable when user is a point. Representing the user as a hypercuboid can effectively address this problem. }
\label{fig:hypercubioid}
\end{center}
\end{figure}

To achieve this, a composition points-to-hypercuboid distance is adopted. It consists of an outside distance and an inside distance. Firstly, we find the nearest point on the surface of the hypercuboid to the item, which is performed with the following expression:
\begin{equation}
    \mathbf{p}_{ui} = \min(\mathbf{p}_u^{\max}, \max (\mathbf{p}_u^{\min}, \mathbf{V}_i)),
\end{equation}
where $\mathbf{p}_u^{\max}$ and $\mathbf{p}_u^{\min}$ denote the upper-right corner and lower-left corner of the user hypercuboid (shown in Figure \ref{fig:hypercubioid}). Their formal definitions are:
\begin{align*}
\begin{split}
    \mathbf{p}_u^{\max} &= \mathbf{C}_u + \mathbf{F}_u \\
    \mathbf{p}_u^{\min} &= \mathbf{C}_u - \mathbf{F}_u.
\end{split}
\end{align*}

It is clear that $\mathbf{p}_{ui}$ is relevant to both the user hypercuboid and the item embedding. The movement of either will result in transformation of this point. 

Then, the outside distance is the distance between $\mathbf{p}_{ui}$ and $\mathbf{V}_i$, and the inside distance indicates the distance between $\mathbf{p}_{ui}$ and the center of the hypercuboid. So we have:
\begin{align*}
\begin{split}
      \ell_{\text{out}}(u,i) &= \|\mathbf{p}_{ui}- \mathbf{V}_i \|_{2}^{2}\\
    \ell_{\text{in}}(u,i) &= \|\mathbf{p}_{ui} - \mathbf{C}_u\|^{2}_{2},
\end{split}
\end{align*}
where euclidean distance is used. Then, we compose these two distances as follows to measure the user-item relationships.
\begin{equation}
 \ell (u, i) =  \ell_{\text{out}}(u,i) +  \gamma \cdot   \ell_{\text{in}}(u,i),
 \label{final1}
\end{equation}
where the coefficient $\gamma$ is employed to control the contribution of the inside distance. Theoretically, if we set $\gamma=0$, we presume that the user is interested in an item regardless of its exact coordinate as long as this item is located in the user hypercuboid. This might alleviate the congestive problem of common approaches which map users and her interacted items onto the same points~\cite{tay2018latent}.

This paradigm may suggest two benefits.
On one hand, a user hypercuboid can be regarded as a set that contains countless item points, which means that it could enwrap as many items as the user likes without much loss of information, enabling higher capacities.
On the other hand, the composition distance in \eqref{final1} could overcome an obvious impediment of the direct distance adopted in the existing body of literature. 
To explain the latter,
suppose that we have two items located at the circumference of the circle with center $\mathbf{C}_u$. With commonly used euclidean distance, the system will conclude that this user likes both items equally. 
However, consider the case in Figure \ref{fig:hypercubioid}.
As we can see, this might not be true when the edges of the hypercuboid is taken into account: it is clear that the distances to the hypercuboid of $\text{item}_1$ and $\text{item}_2$ are not equal. This is provable with the cosine theorem, or by directly setting the hyper-parameter $\gamma$ to $0$. In this example, we can clearly conclude that $\ell(u,\text{item}_2) > \ell(u, \text{item}_1)$ so this user is more interested in the $\text{item}_1$.
One might think of the offset as a range for user preference in each dimension. If the $x$-axis indicates prices and the $z$-axis represents styles,  it is evident that this user has less tolerance in style than that in the price factor. As such, even though $\text{item}_2$ has a more acceptable price than $\text{item}_1$ for this user, she still likes $\text{item}_1$ more because the $\text{item}_2$ does not suit her taste in style at all.

To ensure the items in a hypercuboid to be higher ranked than items outside the hypercuboid, we can add an additional distance to $\ell(u, i)$,
\begin{equation}
  \ell_{\text{add}}(u,i) = 2 (\sigma (\alpha \parallel \mathbf{p}_{ui} - \mathbf{V}_i \parallel^2_2) - \frac{1}{2} )  \parallel \mathbf{F}_u \parallel^2_2,
\end{equation}
where $\alpha$ is set to a value greater than $100$ and $\sigma$ is the \textit{sigmoid} function. This distance equals to zero when item $i$ is in the hypercuboid. Otherwise, it is approximately greater than the half of the maximum diagonal length of the hypercuboid. 

% \subsection{Additional Distance}
% Ideally, we aim to use a hypercuboid to accommodate the items the user likes. Nevertheless, the distance function (\ref{final1}) cannot guarantee this property unless we set $\lambda$ to zero. However, setting it to zero will weaken the capability in ranking the items in the hypercuboid. To remit this issue, we proposed an additional distance:
% \begin{equation}
%   \ell_{\text{add}}(u,i) = 2 (\sigma (\alpha \parallel \mathbf{p}_{ui} - \mathbf{V}_i \parallel^2_2) - \frac{1}{2} )  \parallel \mathbf{F}_u \parallel^2_2,
% \end{equation}
% where $\alpha$ is set to a value greater than $100$ and $\sigma$ is the sigmoid function. This distance will equal to zero when item $i$ is in the hypercuboid. Otherwise, it is approximately greater than the half of the maximum diagonal length of the hypercuboid. 

% We add this additional distance to $\ell(u, i)$, which ensures the items in hypercuboid to be higher ranked than other items on the whole.

\subsection{Variants of Hypercuboids}
\begin{figure}[t]
\begin{center}
\begin{minipage}[r]{3.5cm}
\includegraphics[width=3.5cm]{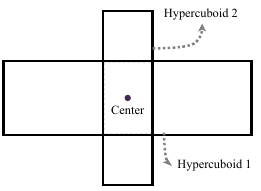}
\centering{(a) Concentric Hypercuboids}
\end{minipage}
\hspace{4em}
\begin{minipage}[r]{3.5cm}
\includegraphics[width=3.5cm]{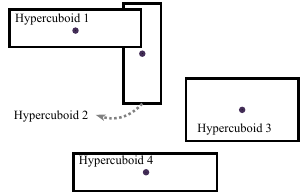}
\centering{(b) Multiple Independent Hypercuboids}
\end{minipage}
\caption{Two variants of hypercuboid.}
\label{fig:multiple}
\end{center}
\end{figure}

To further enhance the capability, we devise the following two variants of hypercuboid. The rational is to generate multiple hypercuboids, with each one providing a different view for user interests. This could improve the capability in capturing the diversities of user interests. Figure \ref{fig:multiple} illustrates these two variants.

\subsubsection{Concentric Hypercuboids}

One solution is to use multiple concentric hypercuboids to represent each user. In this case, we add several offset embeddings and keep the center unchanged. With concentric hypercuboid we can not only extend the regular hypercuboid into irregular space, but also mitigate the above problem. Suppose we have $M$ offsets, there are two distances for each hypercuboid. We integrate them into the following final distance:
\begin{equation}
\ell(u, i) =  \sum_{j=1}^M \ell_{\text{out}}^j(u,i) +  \gamma \cdot  \min( \ell_{\text{in}}^1(u,i), ...,  \ell_{\text{in}}^M(u,i)).
\label{concentric}
\end{equation}

All the outside distances are summed up while only the minimum inside distance is kept. The motivation is to weaken the effects of the inside distances.

\subsubsection{Multiple Independent Hypercuboids} To further enhance the capability in modeling diverse interests, we can also use multiple hypercuboids with different centers and offsets. The final distance function is defined as:
\begin{equation}
\ell(u, i) =  \min (\ell^1(u, i), ..., \ell^M(u,i)). 
\label{multiple}
\end{equation}

As for the addition distance, we can take the maximum additional distance from all hypercuboids.

\subsection{Learning the Centers and Offsets}
An important step is to learn the center and offset of the user hypercuboid. Generally, user interests can be derived from users' historical behaviors, e.g., buy, watch, and click. So it is natural to learn the centers and offsets from the past interaction log. In most practical applications, user actions are often ordered and timestamped, which also records users' interests evolution. Formally, we make use of the last $L$ actions of users, and aim to predict the next recommendable item(s).  At the embedding layer, we get an embedding matrix $\mathbf{S}_{u}^{(t)} \in \mathbb{R}^{L \times d}$ , to represent these $L$ items at time-step $t$,  by looking up the item embedding matrix $\mathbf{V}$ with the index of the $L$ items.

To capture the shifts of user interests, we employ a stacked neural network consisting of a bidirectional LSTM~\cite{schuster1997bidirectional} and a self-attention module over the $L$ sequential actions. 
\begin{equation}
    \mathbf{S}' = \text{Bi-LSTM}(\mathbf{S}_{u}^{(t)}).
\end{equation}

The bidirectional LSTM is used to enhance the sequence representation capability. Then, a self-attention network is applied.
\begin{equation}
    \mathbf{a}_{u}^{(t)} = \text{softmax}(\frac{f(\mathbf{S}') \cdot f(\mathbf{S}')^\top}{\sqrt{d}}) \cdot \mathbf{S}_{u}^{(t)},
\end{equation}
where $\mathbf{a} \in \mathbb{R}^{L \times d}$ is transformed representation for the current sequence; $f$ is nonlinear layer. It is then passed into a pooling layer to get a $d$ dimensional vector.
\begin{equation}
    \mathbf{s}_{u}^{(t)} = \text{pooling}(\mathbf{a}_{u}^{(t)}),
\end{equation}
where the pooling function can be average-pooling, max-pooling, min-pooling, or sum-pooling. We can use $\mathbf{s}_{u}^{(t)}$ to represent the center, the offset, or both.

In addition, we notice that the offset of hypercuboid usually needs to memorize mass amount of information. For example, to get the price interval, we need to store the prices of all items the user bought to accurately infer her acceptable price range. To this end, we adopt a key-value memory network to perform the memorization process.  Let $\mathbf{M} \in \mathbb{R}^{d \times N}$ denote the memory and $\mathbf{K} \in \mathbb{R}^{d \times N}$ denote the key matrix, where $N$ controls the capacity of the memory. 

\begin{figure}[t]
\centering
\includegraphics[height=0.85\linewidth]{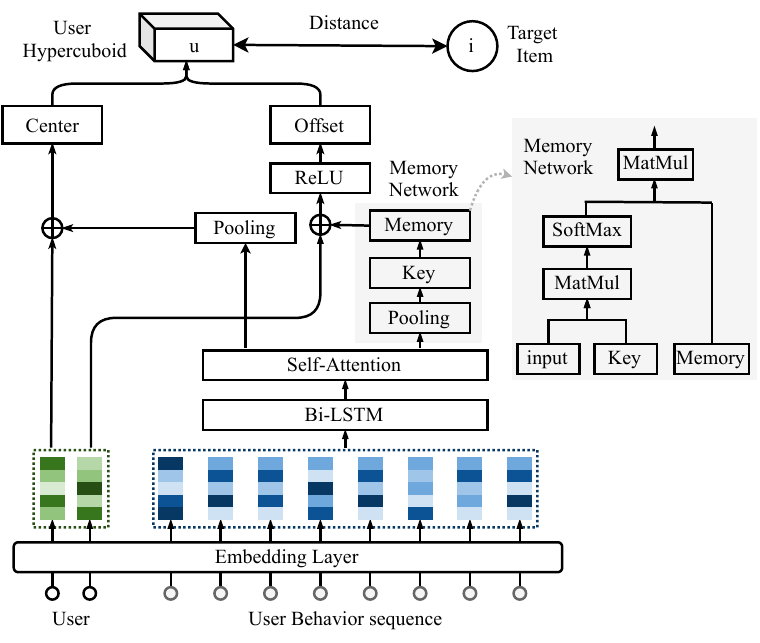}
\centering\caption{The architecture of the proposed model with one hypercuboid for each user.}
\label{fig:arch}
\end{figure}

The input of this memory module is $\mathbf{s}_{u}^{(t)}$, which is used to retrieve the key from the key matrix.
\begin{equation}
    \mathbf{k} = \text{softmax}(\mathbf{s}_{u}^{(t)} \mathbf{K}).
\end{equation}

With this attentive key vector $\mathbf{k}$, we select relevant pieces of information from the memory.
\begin{equation}
    \mathbf{m} = \mathbf{k} \cdot \mathbf{M}^\top.
\end{equation}

We wrap all the offsets with the $ReLU$ function~\cite{goodfellow2016deep} to ensure all the values are non-negative. Figure \ref{fig:arch} illustrates the overall architecture of the proposed model.

For multiple hypercuboids, we can add dense layers over the center/offset and get multiple centers and offsets.

\subsection{Optimization}
In this section, we introduce the objective function and training scheme. We adopt the pairwise ranking loss for the optimizing objective, which is defined as.
\begin{equation}
    \mathcal{L} = \sum_{(u, i) \in \mathcal{T}^+} \sum_{ (u, j)\in \mathcal{T}^-} \text{max}(0, \ell(u,i) + \lambda - \ell(u,j)),
    \label{loss}
\end{equation}
where $\mathcal{T}^+$ and $\mathcal{T}^-$ are the positive user-item pairs and negative user-item pairs respectively, $\lambda$ is the margin that separates the two items. For regularization, standard $L_2$ normalization and dropout~\cite{srivastava2014dropout} are viable.

We use the same negative sampling strategy as that of HGN~\cite{ma2019hierarchical}. Our model is end-to-end differentiable and an adaptive gradient descent optimizer is used. It is noteworthy that other losses such as Bayesian personalized ranking loss or cross entropy loss are also applicable.

\section{Experiments}
In this section, we present our experimental setup and empirical evaluation to verify the effectiveness of our method.

\begin{table}[]
\caption{Statistics of all datasets used in our experiment.}
\begin{tabular}{|c|c|c|c|c|}
\hline
\textbf{Datasets}  & \textbf{Interactions} & \# \textbf{Users} & \# \textbf{Item}   & \% \textbf{Density}\\ \hline
Books      &      1,856,747   &  52,406      &   41,264      &  0.086     \\ \hline
CDs         &     475,265    &  17,052      &   35,118    &   0.079    \\ \hline
Movie \& Tvs        & 694,970        &    23,337   &  33,582       &    0.089   \\ \hline
E-commerce         & 5,560,954      &    104,929    &    69,563     &   0.076    \\ \hline

\begin{tabular}[c]{@{}c@{}}Alibaba \\  Dataset V1\end{tabular} &    28,926,924     &   145,057     &    92,364     &   0.216    \\ \hline
\begin{tabular}[c]{@{}c@{}}Alibaba \\  Dataset V2\end{tabular} &   8,558,837      &   131,692     &    109,670     &    0.059   \\ \hline
\end{tabular}

\label{exp:dataset}
\end{table}

\subsection{Datasets} We conduct our empirical evaluation based on four public well-studied benchmark datasets and two commercial datasets. The statistics of the datasets are reported in Table \ref{exp:dataset}. 

\subsubsection{Public Datasets} Four public datasets from different domains are used, including Amazon-Books, Amazon-Movies\&TVs, Amazon-CDs~\cite{he2016ups,ma2019hierarchical,mcauley2015image}  and an E-commerce dataset~\cite{lv2019sdm}. The feedback in the Amazon datasets is ratings from 1 to 5, we treated those with no less than 4 as positive feedback. For the E-commerce dataset, we use the version provided by Lv et al.~\cite{lv2019sdm} which consists of user-item interaction logs of four weeks.  To filter noisy data, we discard users with fewer than ten actions and items with fewer than 5 associated actions for each dataset~\cite{ma2019hierarchical}. 

\subsubsection{Commercial Datasets} This dataset is from an online retailer platform, Alibaba.  It is collected from the interaction logs generated in seven days. We sampled two versions for two groups of users. The first version consists of users with 100 to 500 interactions and the second version consists of users with 50 to 100 interactions.

We target at the next item prediction task where the aim is to predict a user's next interaction. As such, each dataset is split into three subsets based on the chronological order of interactions, with $70\%$ of the user sequence as the training set, the next $10\%$ as the validation set and the latest $20\%$ as the test set. Same as~\cite{ma2019hierarchical}, the validation set is used during the inference stage.

\subsection{Evaluation Measures}

Three commonplace evaluation measures~\cite{shani2011evaluating} including Recall@k, Normalized Discounted Cumulative Gain (NDCG@k), and Mean Average Precision (MAP@k) are adopted. For each user, we rank \textbf{all} the items (exclude items the user has rated) and select the highest ranked k items for recommendation. Recall at $k$ indicates the proportion of relevant items found in the top-$k$ recommendations. NDCG at $k$ is a measure of ranking quality and the position of correctly recommended item is taken into account.  MAP at $k$ is the mean of the average precision scores of all users and is a precision measure that gives larger credit to correctly recommended items. Definitions are omitted for brevity.

%https://towardsdatascience.com/evaluate-your-recommendation-engine-using-ndcg-759a851452d1

\subsection{Compared Methods} 
We compare our method with two groups of well-established and competitive baselines.
The first group contains two non-neural recommendation methods.
\begin{itemize}
    \item \textbf{Matrix Factorization with Bayesian Personalized Ranking (BPRMF)} ~\cite{10.5555/1795114.1795167} is a classical method with pairwise Bayesian ranking loss. It models user-item interactions with matrix factorization.
    
    \item \textbf{Translational Recommender Systems (TransRec)}~\cite{he2017translation,10.1145/3109859.3109882} is a sequential recommendation model where users and items are embedded into a transition space following the $\textit{prev. item} + \textit{user} \approx \textit{next item}$ translation operation.
\end{itemize}
Five neural networks based recommendation models are included in the second group.
\begin{itemize}
     \item \textbf{YouTubeDNN}~\cite{covington2016deep} is a deep neural networks based recommendation model proposed by YouTube~\footnote{https://www.youtube.com/}. User vectors, vectors of the latest interacted items and the target item are concatenated and are fed into multiple dense layers.
    \item \textbf{Gated Recurrent Unit for Recommendation (GRU4Rec)}~\cite{hidasi2015session} is a recurrent neural network based model for session-based recommendation (a user is a session).
    \item \textbf{Convolutional Sequence Embedding Recommendation Model (Caser)}~\cite{tang2018personalized} is a convolutional neural networks based sequence-aware recommendation model. The proposed framework captures the sequential patterns at both point-level and union-level.
    \item \textbf{Self-Attentive Sequential Recommendation (SASRec)}~\cite{kang2018self} is a self-attention~\cite{vaswani2017attention} based recommendation model, where user sequences are modelled with a transformer like network. 
    \item \textbf{Hierarchical Gating Networks (HGN)}~\cite{ma2019hierarchical} is a state of the art
 sequential recommendation model which is based on a novel gating mechanism and includes item-item relations. 
\end{itemize}

\subsection{Implementation Details}
We implemented our model and baselines with tensorflow~\footnote{https://www.tensorflow.org/}. For all models,  the embedding size $d$ is set to $100$ for fair comparison, the sequence length $L$ and the target length $T$ are set to 5 and 3, respectively. Adagrad optimizer~\cite{duchi2011adaptive} is used for model parameters updates and the learning rate is set to $0.05$. The $\ell_2$ regularization rate is tuned amongst $\{0.1, 0.01, 0.001, 1\text{e-}4, 1\text{e-}6\}$. The margin of hinge loss $\lambda$ is set to $0.5$. The coefficient $\gamma$ is selected from $\{0.1, 0.3, 0.5, 0.7, 0.9\}$. Learnable parameters are initialized with a normal distribution and trained from scratch. The batch size is set to $4,096$. The memory capacity $N$ is determined from $\{10, 20, 30, 40, 50\}$.  For all pooling operations in the model, average pooling is used by default. All the hyper-parameters are tuned with the validation set. Experiments are executed for five times and the average performances are reported.

% Precision at k is the proportion of recommended items in the top-k set that are relevant

% Please add the following required packages to your document preamble:
% \usepackage{multirow}
% Please add the following required packages to your document preamble:
% \usepackage{multirow}
\begin{table*}[t]
\caption{Performance comparison on two Alibaba datasets. Only k=10 and k=50 are reported due to space limitation. Only one hypercuboid is used in our model.}
\begin{tabular}{|c|c|c|c|c|c|c|c|c|c|c|}
\hline
Datasets                                                                       & Measure                 & k  & BPRMF  & TransRec & YouTubeDNN & GRU4Rec & Caser  & SASRec & HGN                         & Ours   \\ \hline
\multirow{6}{*}{\begin{tabular}[c]{@{}c@{}}Alibaba \\ Dataset V1\end{tabular}} & \multirow{2}{*}{Recall} & 10 & 0.0156 & 0.0192   & 0.0251     & 0.0196  & 0.0227 & 0.0233 & 0.0271                      & \textbf{0.0299} \\ \cline{3-11} 
                                                                               &                         & 50 & 0.0459 & 0.0608   & 0.0776     & 0.0631  & 0.0683 & 0.0692 & 0.0774                      & \textbf{0.0885} \\ \cline{2-11} 
                                                                               & \multirow{2}{*}{MAP}    & 10 & 0.0284 & 0.03.0   & 0.0487     & 0.0367  & 0.0451 & 0.0466 & 0.0558                      & \textbf{0.0610} \\ \cline{3-11} 
                                                                               &                         & 50 & 0.0108 & 0.0151   & 0.0211     & 0.0163  & 0.0192 & 0.0192 & \textbf{0.0232}                      & 0.0231 \\ \cline{2-11} 
                                                                               & \multirow{2}{*}{NDCG}   & 10 & 0.0665 & 0.0789   & 0.1041     & 0.0808  & 0.0952 & 0.0989 & \multicolumn{1}{l|}{0.1141} & \textbf{0.1255} \\ \cline{3-11} 
                                                                               &                         & 50 & 0.0427 & 0.0543   & 0.0702     & 0.0562  & 0.0627 & 0.0641 & 0.0724                      & \textbf{0.0814} \\ \hline
\multirow{6}{*}{\begin{tabular}[c]{@{}c@{}}Alibaba\\ Dataset V2\end{tabular}}  & \multirow{2}{*}{Recall} & 10 & 0.0370 & 0.0465   & 0.0622     & 0.0344  & 0.0639 & 0.0619 & 0.0649                      & \textbf{0.0667} \\ \cline{3-11} 
                                                                               &                         & 50 & 0.0997 & 0.0128  & 0.1655     & 0.1069  & 0.1664 & 0.1568 & 0.1626                      & \textbf{0.1764} \\ \cline{2-11} 
                                                                               & \multirow{2}{*}{MAP}    & 10 & 0.0223 & 0.0273   & 0.0396     & 0.0186  & 0.0415 & 0.0413 & 0.0425                      & \textbf{0.0431} \\ \cline{3-11} 
                                                                               &                         & 50 & 0.0233 & 0.0303   & 0.0441     & 0.0218  & 0.0451 & 0.0441 & 0.0454                      & \textbf{0.0470} \\ \cline{2-11} 
                                                                               & \multirow{2}{*}{NDCG}   & 10 & 0.0536 & 0.0648   & 0.0881     & 0.0470  & 0.0913 & 0.0897 & \multicolumn{1}{l|}{0.0930} & \textbf{0.0947} \\ \cline{3-11} 
                                                                               &                         & 50 & 0.0323 & 0.0405   & 0.0537     & 0.0320  & 0.0543 & 0.0521 & 0.0540                      & \textbf{0.0569} \\ \hline
\end{tabular}
\label{allresults}
\end{table*}
% \begin{table*}[]
% \caption{NDCG@10 on six benchmark datasets. Only one hypercuboid is used in our model. The best performance is in boldface and the second best is underlined. Numbers are in \%.}
% \begin{tabular}{|c|c|c|c|c|c|c|c|c|}
% \hline
% Datasets & BPRMF & TransRec & YouTubeDNN  & GRU4Rec & Caser  & SASRec & HGN    & \textbf{Ours}  \\ \hline
% % Books  & 1.74 & 1.96   &2.82  & 2.13 & 2.58  & 2.94 &\underline{3.55} & \textbf{3.83} \\ \cline{2-10} \hline
% % CDs &  2.19      &  2.26   &   1.84 & 1.83  &  1.78   &     2.13    &      \underline{2.78}  &   \textbf{ 2.96 }       \\ \cline{2-10}
% %  \hline
% % Movies\&TVs  & 1.53   &2.08  &   1.63  & 1.58 &   1.69     &   1.94     &      \textbf{ 2.32} &   \textbf{ 2.32 }       \\ \cline{2-10}  \hline
% % E-commerce  &  7.48    &   6.99 &  11.38 &6.40  &  8.92      &   6.46     &       \underline{11.60} &     \textbf{12.90  }        \\ \cline{2-10} \hline
  
% \begin{tabular}[c]{@{}c@{}}Company \\  Dataset V1\end{tabular}  &6.65    &   7.89 &  10.41 & 8.08   & 9.52       &   9.89     &   \underline{ 11.41 }   &   \textbf{12.55}         \\ \cline{2-10} \hline
%  \begin{tabular}[c]{@{}c@{}}Company \\  Dataset V2\end{tabular} & 5.36    & 6.48  & 8.81  & 4.70  &  9.13     &  8.97      &    \underline{9.30}    &       \textbf{9.47 }      \\ \cline{2-10} \hline
% \end{tabular}
% \label{allresults}
% \vspace{-1ex}
% \end{table*}

\subsection{Experimental Results}
Table \ref{allresults} and Figure 5 present the recommendation performance of all baselines and our model with a \textbf{single} hypercuboid on the six datasets. The best performer on each row is highlighted in boldface. We observe that our model achieves the best performance across all datasets, outperforming a myriad of complex architectures. Our model consistently performs better than HGN, SASRec, Caser, and TransRec, which are all recent competitive sequence-aware recommendation methods. Notably, we obtain a clear performance gain over the second best baseline. Furthermore, our method performs much better than the inner product model (BPRMF) and the Euclidean distance based model (TransRec), which ascertains the effectiveness of the proposed distance function.

Among all the neural networks based baselines, HGN outperforms self-attention, RNN, and CNN based approaches. The second best neural networks based model often switches between YouTubeDNN and SASRec. We also note that non-neural models such as BPRMF and TransRec achieve very competitive performance on the three Amazon datasets and the E-commerce dataset. 

% This might because that the time interval for data collection of these three datasets is too large (years), which weakens the effects of sequential activities. 

\begin{figure*}[t]
\begin{center}
\caption{Performance comparison on four public datasets with different $k$. Only one hypercuboid is used in our model.}

\begin{minipage}[t]{4.2cm}
\includegraphics[width=4.2cm]{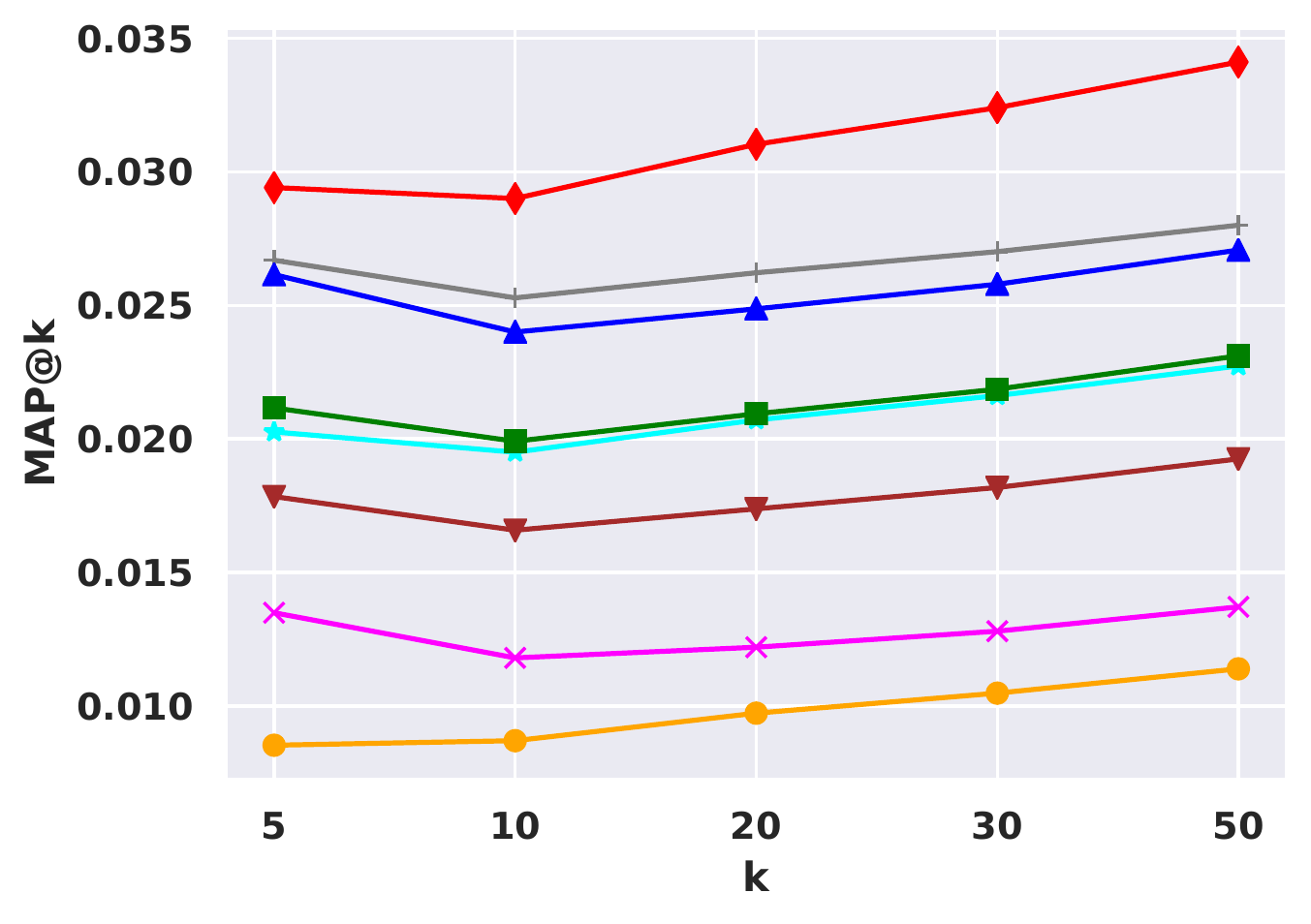}
\centering{MAP@k on Books }
\end{minipage}
\begin{minipage}[t]{4.2cm}
\includegraphics[width=4.2cm]{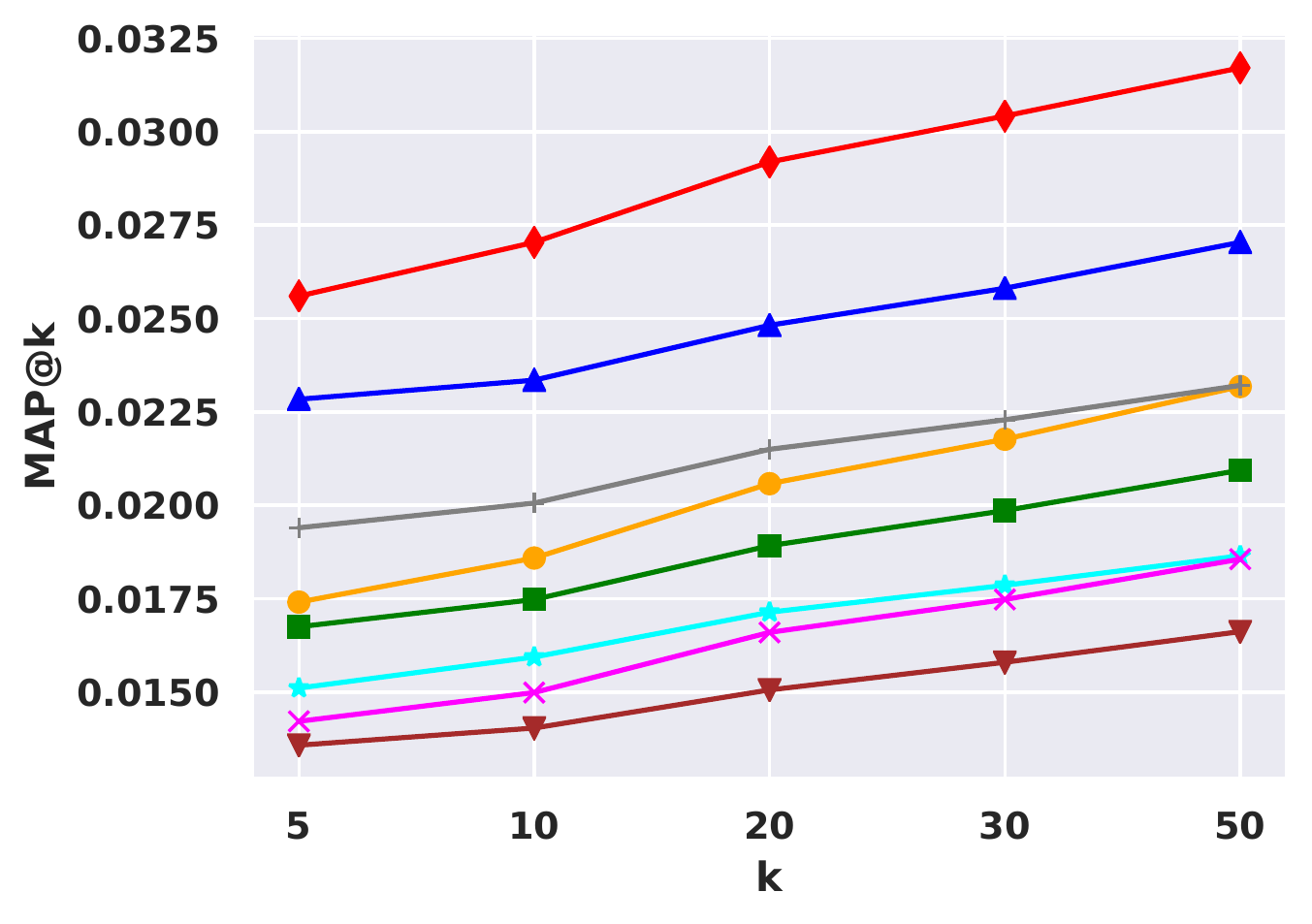}
\centering{MAP@k on CDs }
\end{minipage}
\begin{minipage}[t]{4.2cm}
\includegraphics[width=4.2cm]{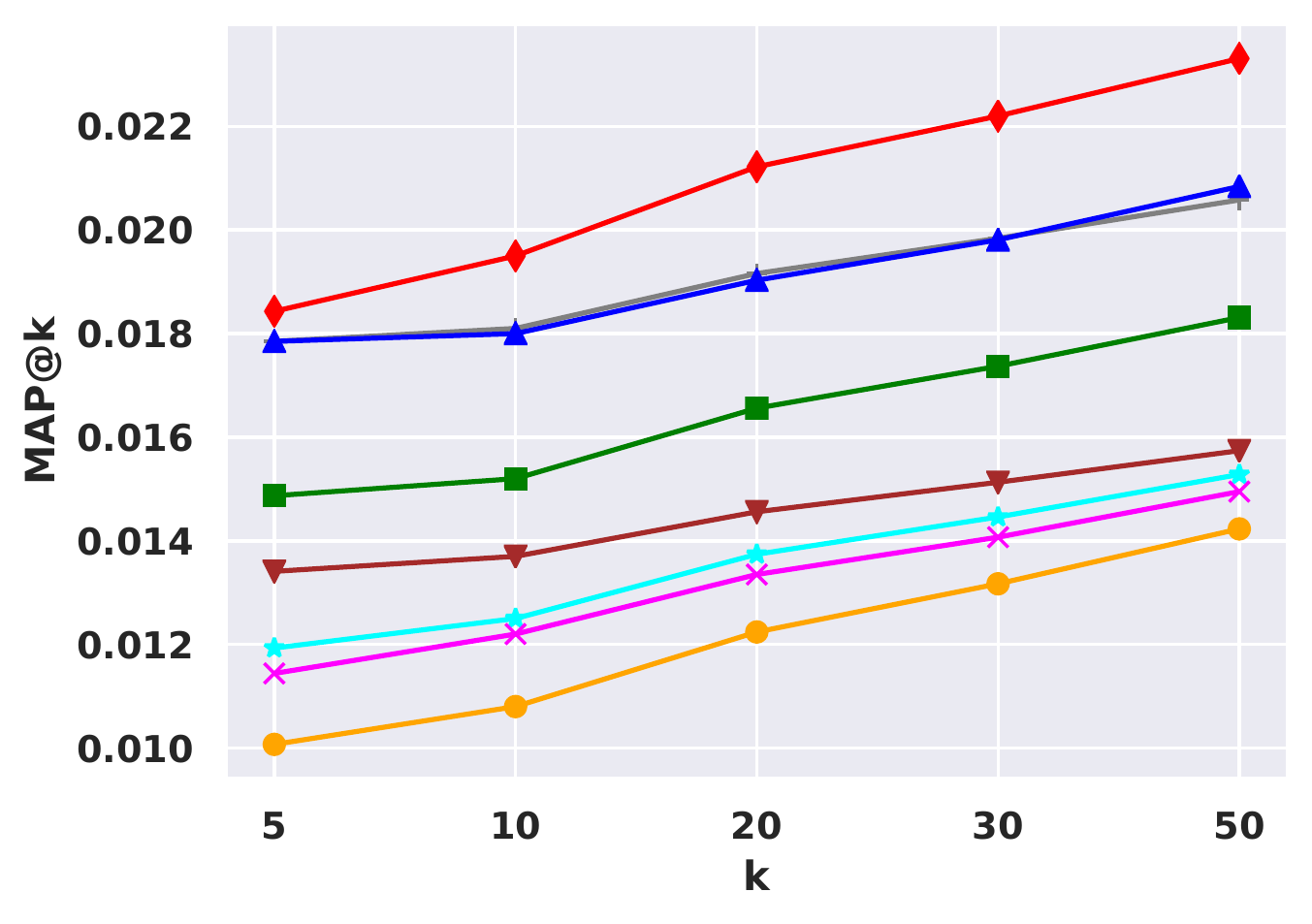}
\centering{MAP@k on Movies\&TVs}
\end{minipage}
\begin{minipage}[t]{4.2cm}
\includegraphics[width=4.2cm]{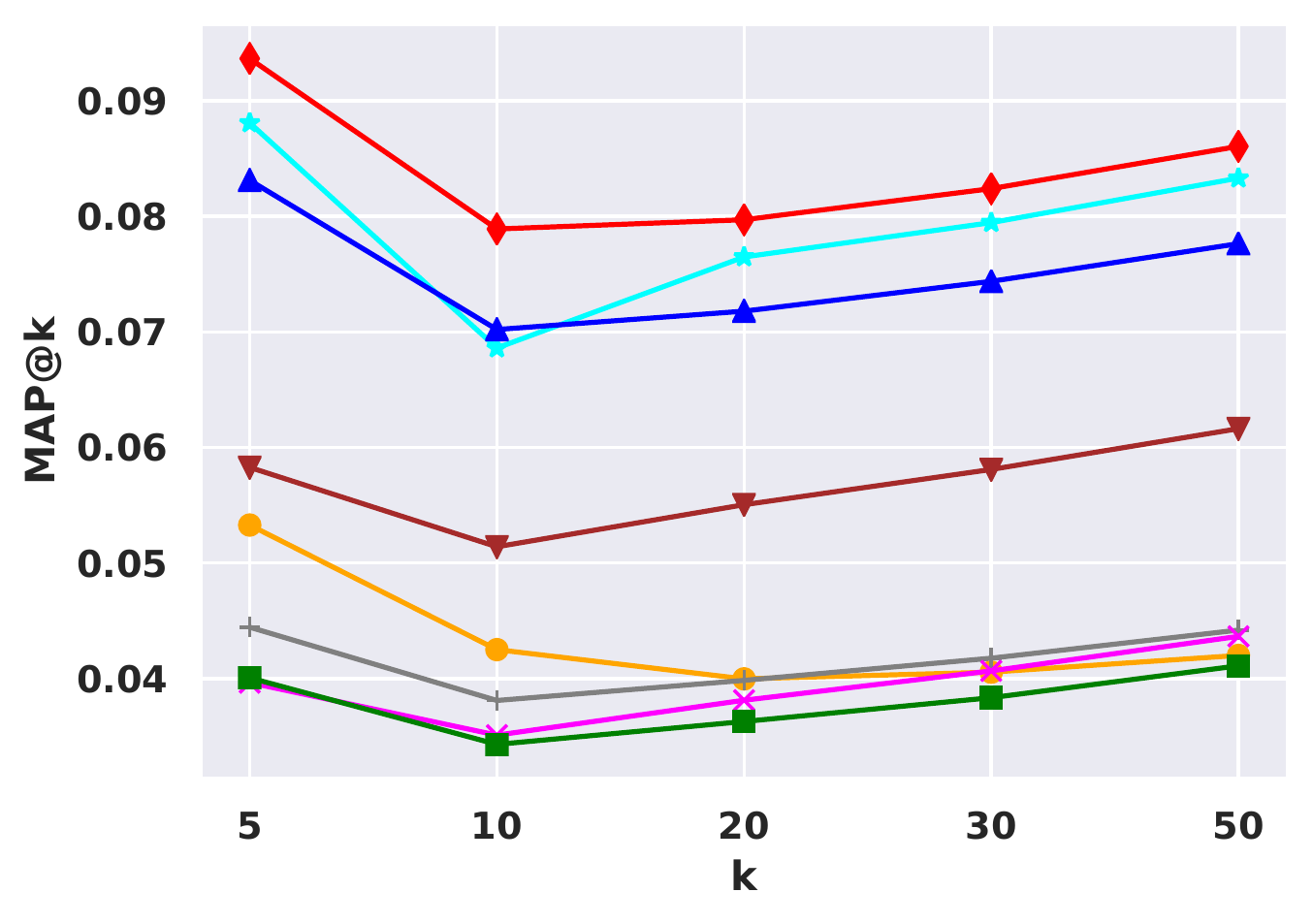}
\centering{MAP@k on E-commerce }
\end{minipage}

\begin{minipage}[t]{4.2cm}
\includegraphics[width=4.2cm]{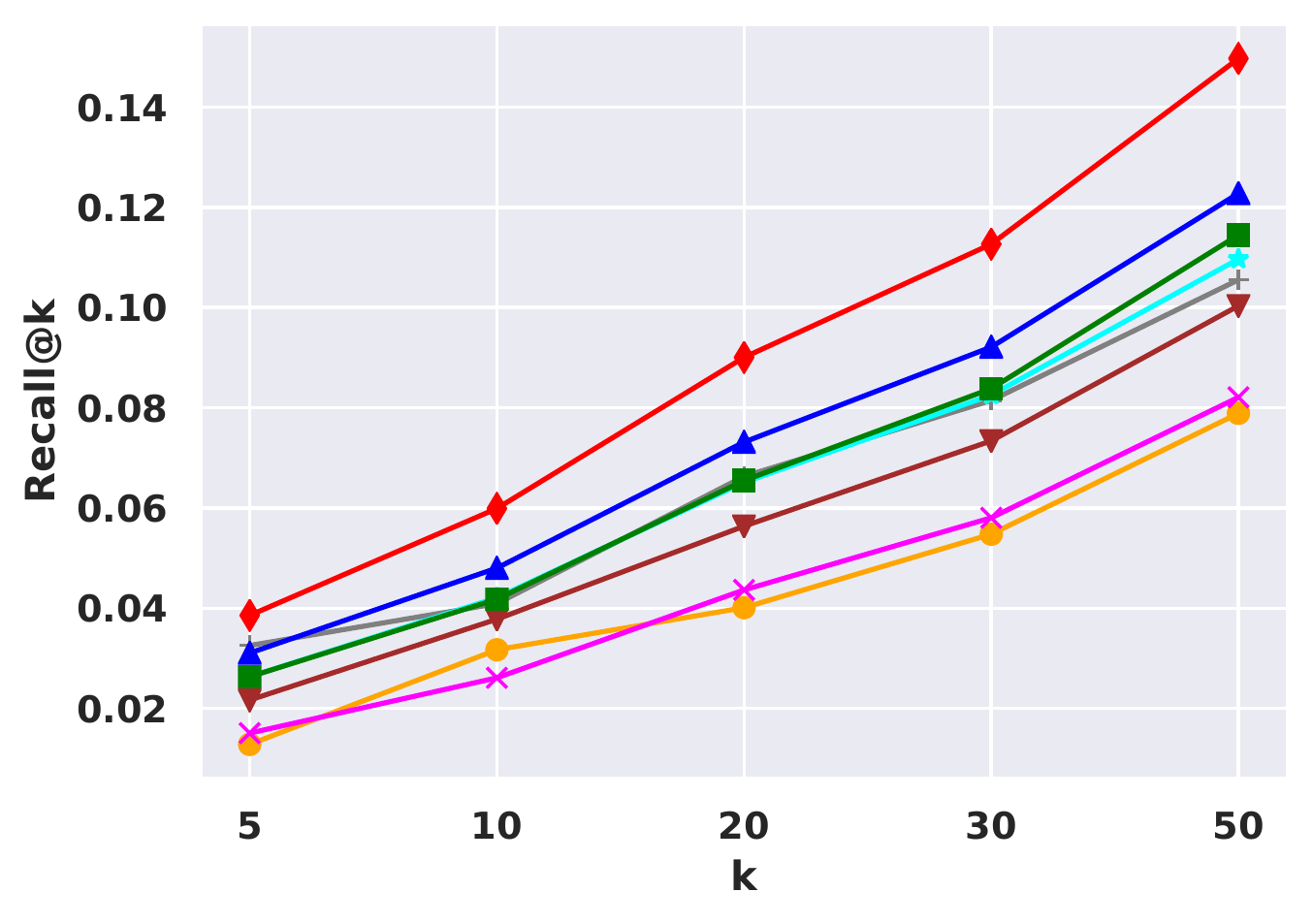}
\centering{Recall@k on Books }
\end{minipage}
\begin{minipage}[t]{4.2cm}
\includegraphics[width=4.2cm]{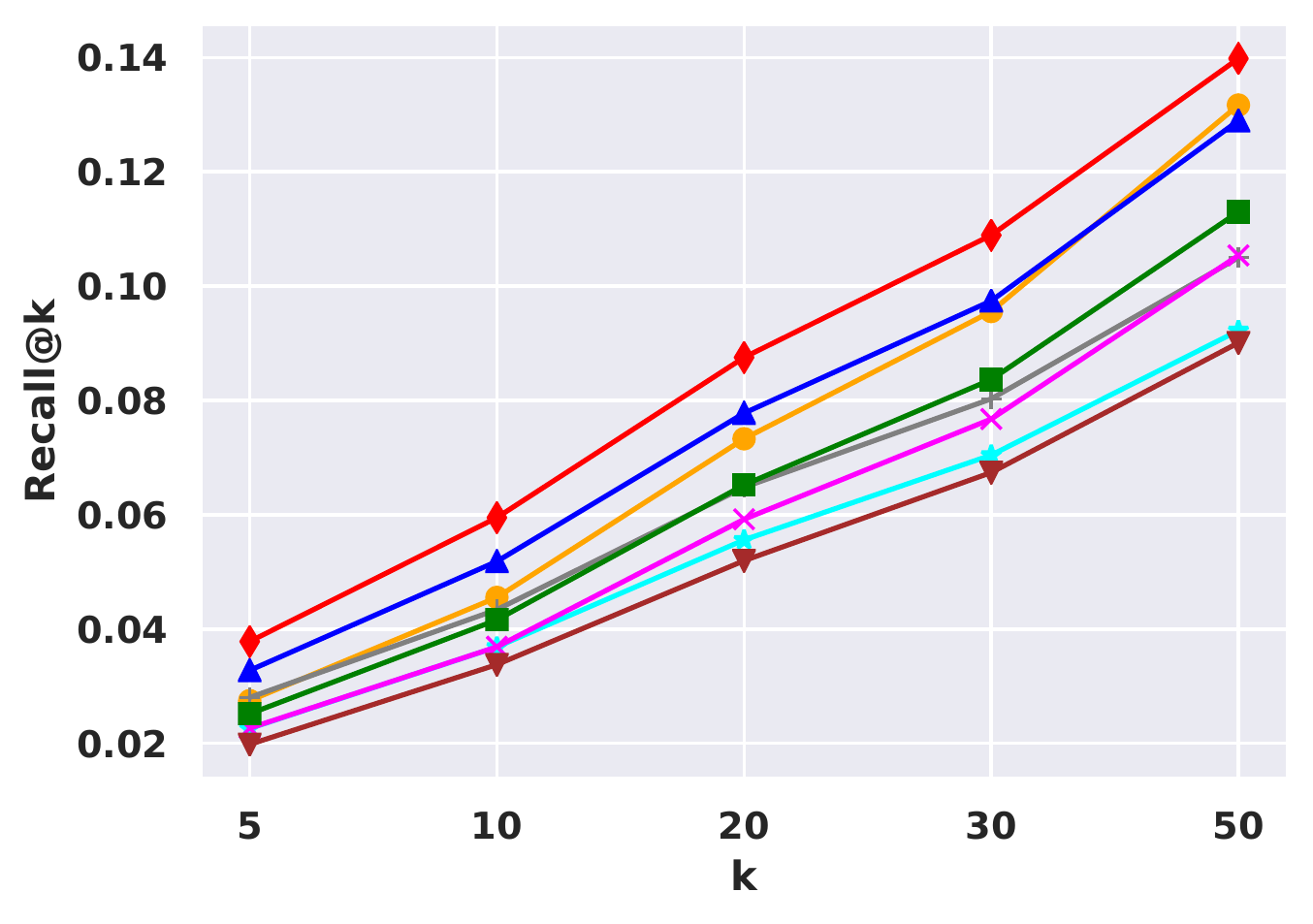}
\centering{Recall@k on CDs}
\end{minipage}
\begin{minipage}[t]{4.2cm}
\includegraphics[width=4.2cm]{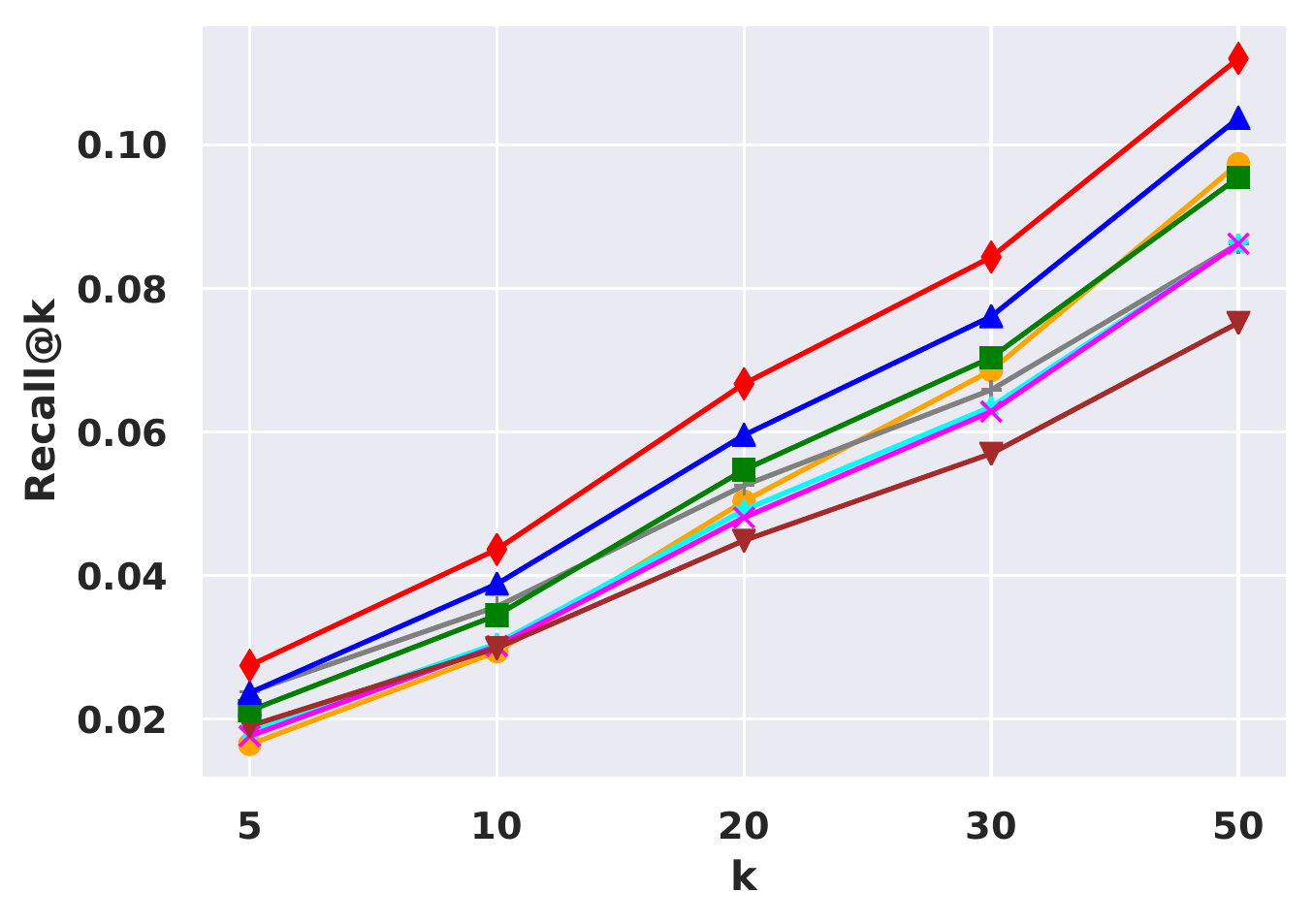}
\centering{Recall@k on Movies\&TVs}
\end{minipage}
\begin{minipage}[t]{4.2cm}
\includegraphics[width=4.2cm]{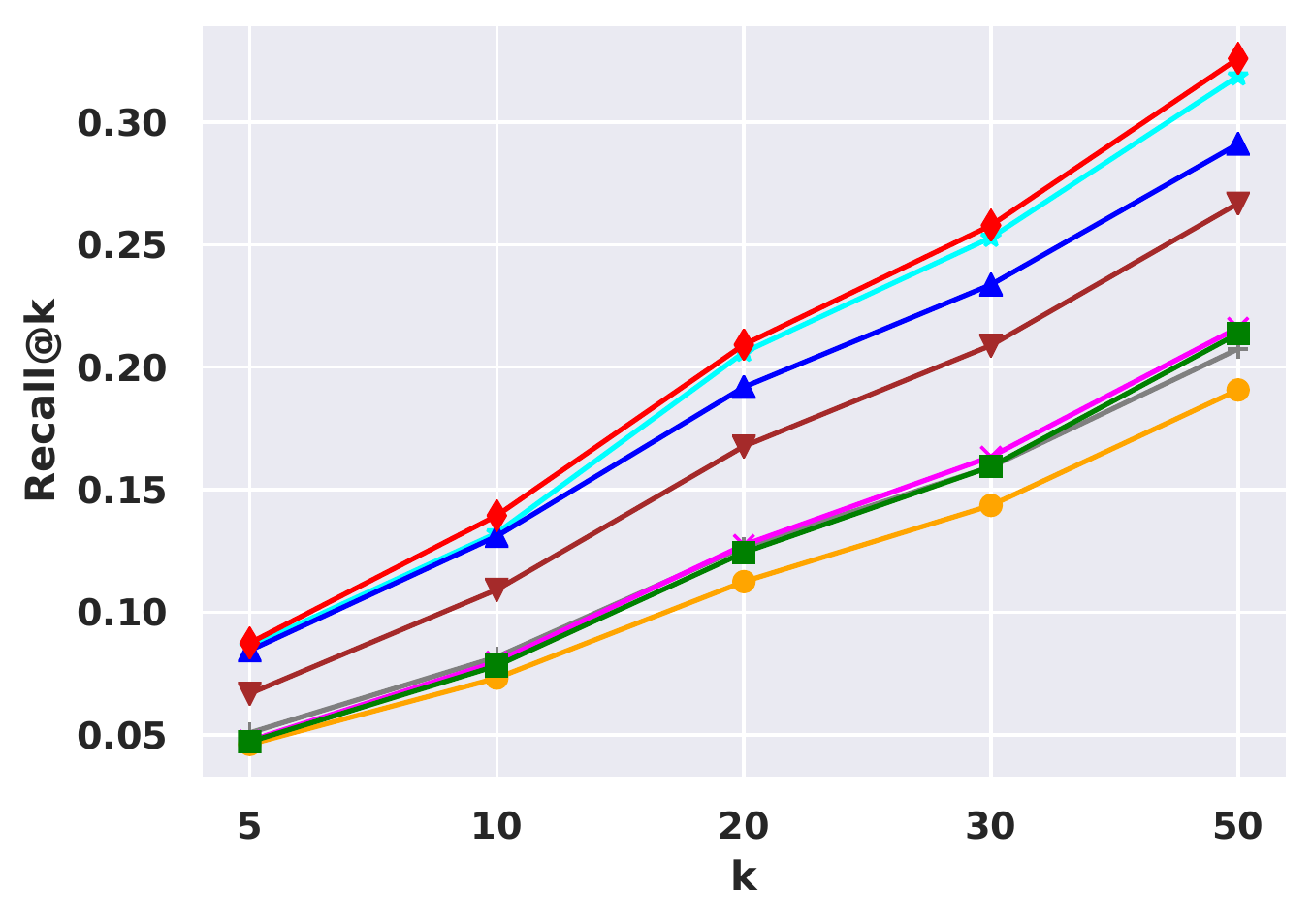}
\centering{Recall@k on E-commerce }
\end{minipage}

\begin{minipage}[t]{4.2cm}
\includegraphics[width=4.2cm]{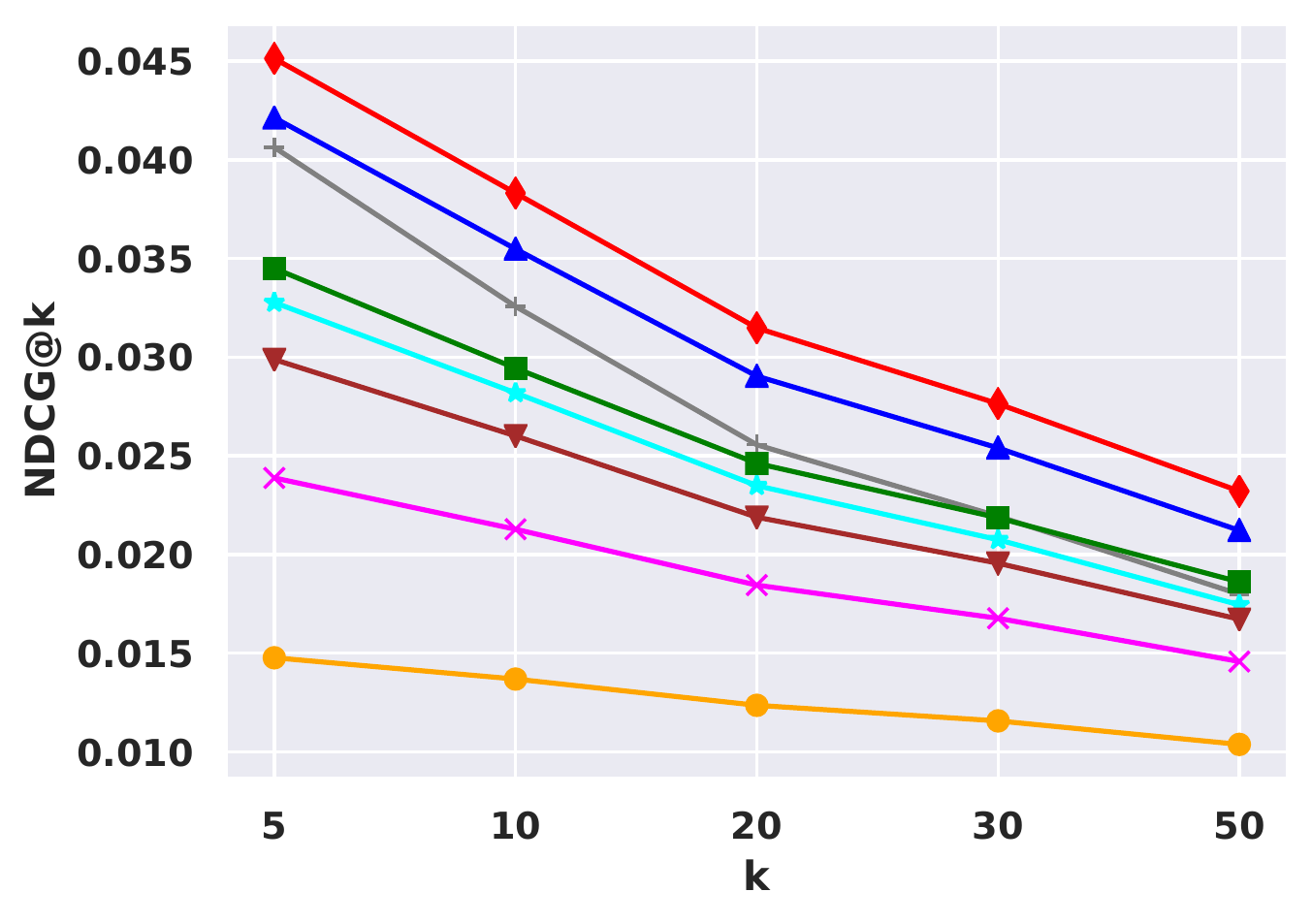}
\centering{NDCG@k on Books}
\end{minipage}
\begin{minipage}[t]{4.2cm}
\includegraphics[width=4.2cm]{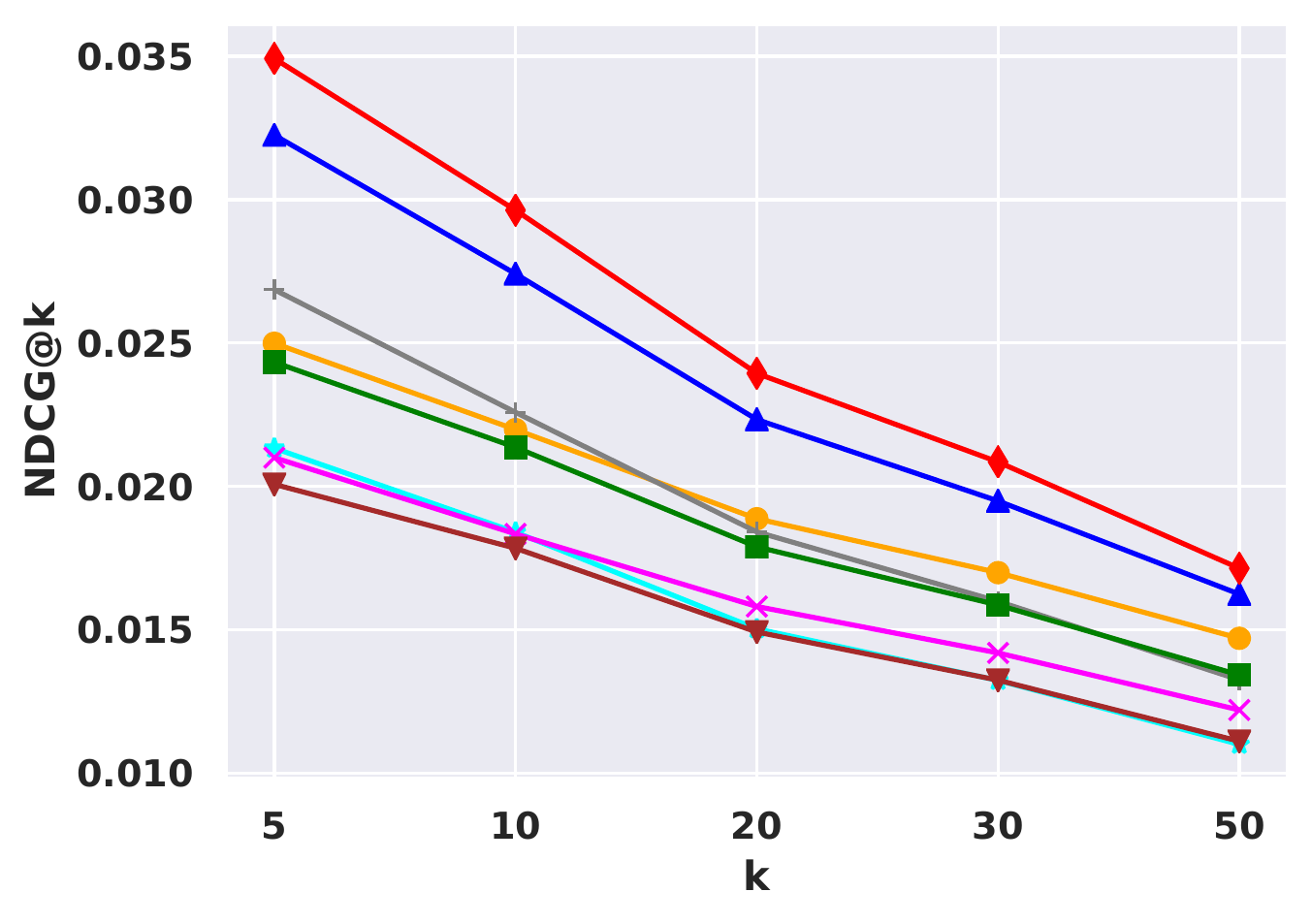}
\centering{NDCG@k on CDs }
\end{minipage}
\begin{minipage}[t]{4.2cm}
\includegraphics[width=4.2cm]{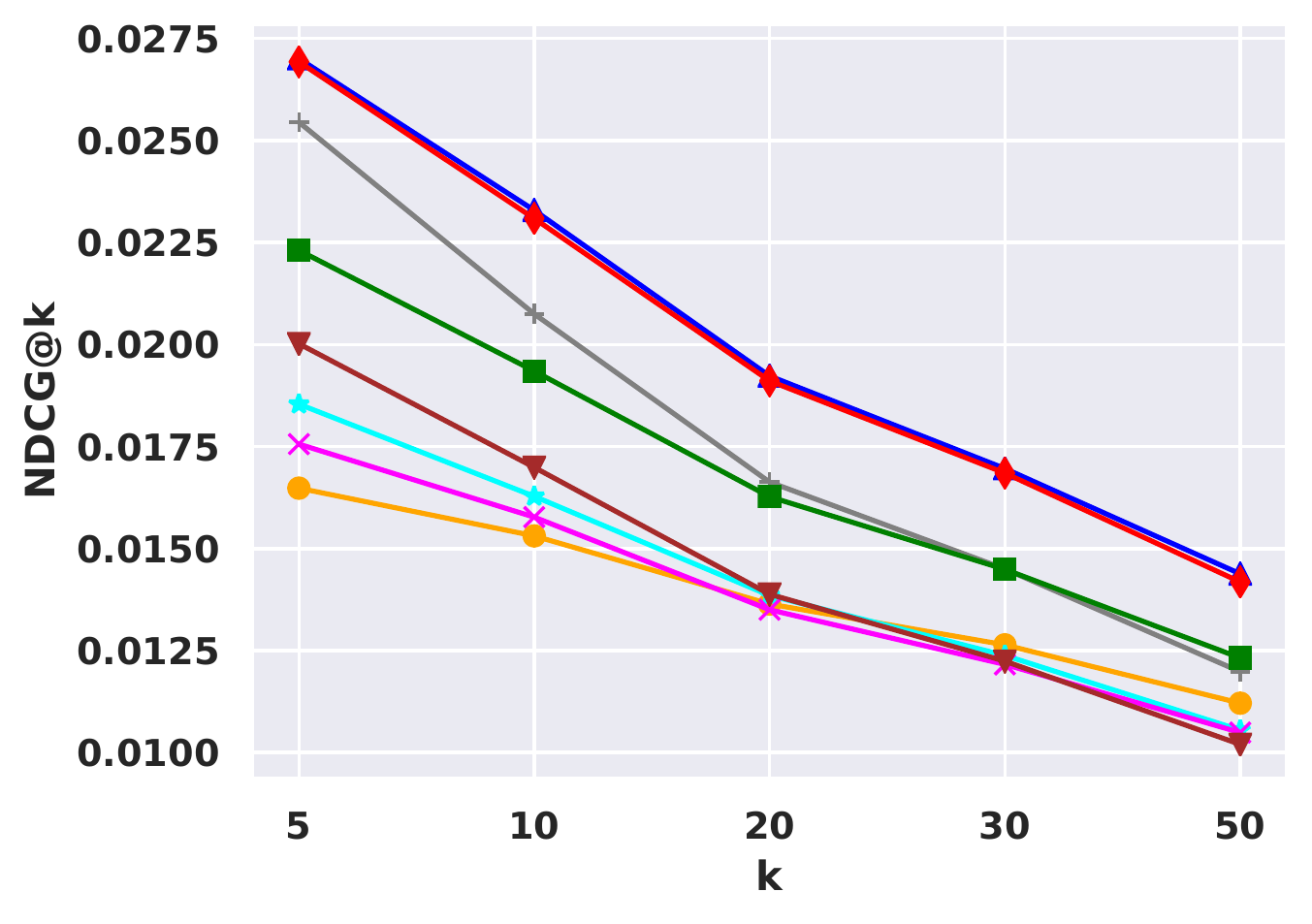}
\centering{NDCG@k on Movies\&TVs}
\end{minipage}
\begin{minipage}[t]{4.2cm}
\includegraphics[width=4.2cm]{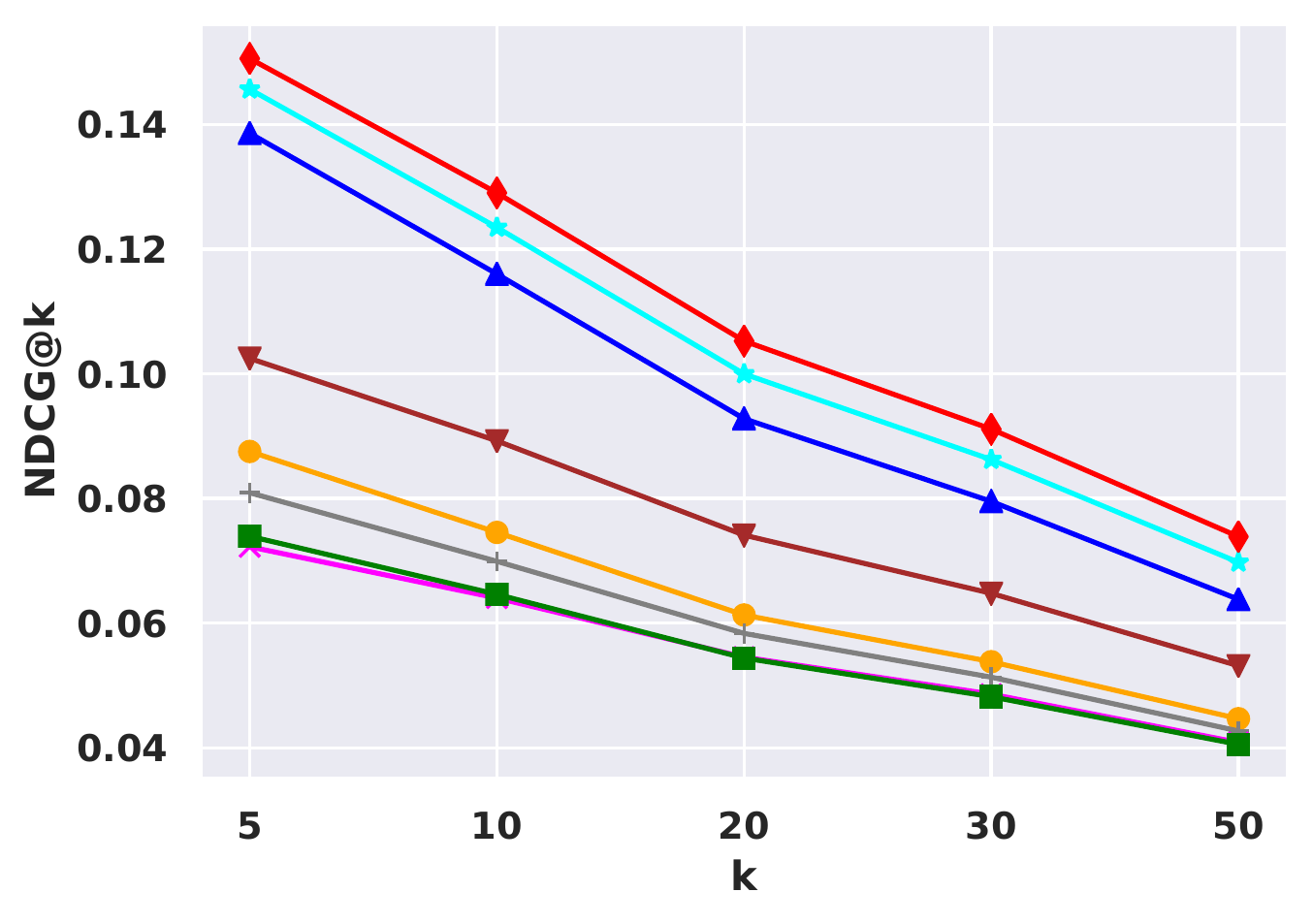}
\centering{NDCG@k on E-commerce}
\end{minipage}

\label{fig:ndcgatk}
\end{center}
\vspace{-2ex}
\end{figure*}

Performance of top 5, 10, 20, 30, 50 on four public datasets are presented in Figure \ref{fig:ndcgatk}. Our model consistently outperforms all the other models, indicating that our model can generate not only more accurate topmost recommendations but also higher quality recommendation lists on the whole.

% \begin{figure}[t]
% \begin{center}
% \caption{Comparison of Recall@k on Books and CDs}
% \begin{minipage}[t]{4.2cm}
% \includegraphics[width=4.2cm]{KDD2020/atk/recall@k.pdf}
% \centering{Recall@k on Books }
% \end{minipage}
% \begin{minipage}[t]{4.2cm}
% \includegraphics[width=4.2cm]{KDD2020/atk/cdsrecall@k.pdf}
% \centering{Recall@k on CDs}
% \end{minipage}

% \label{fig:recallatk}
% \end{center}
% \vspace{-5ex}
% \end{figure}

% \begin{figure}[t]
% \begin{center}
% \caption{Comparison of MAP@k on Movies\&TVs and E-commerce.}

% \begin{minipage}[t]{4.2cm}
% \includegraphics[width=4.2cm]{KDD2020/atk/mtmap@k.pdf}
% \centering{MAP@k on Movies\&TVs }
% \end{minipage}
% \begin{minipage}[t]{4.2cm}
% \includegraphics[width=4.2cm]{KDD2020/atk/jdmap@k.pdf}
% \centering{MAP@k on JD}
% \end{minipage}

% \label{fig:mapatk}
% \end{center}
% \vspace{-4ex}
% \end{figure}

\begin{figure*}[h]
\begin{center}
\begin{minipage}[t]{7cm}
\includegraphics[width=7cm]{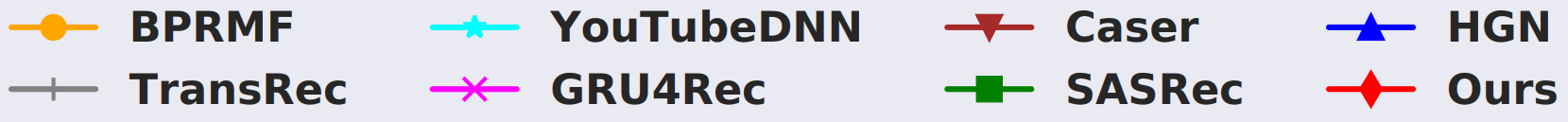}
\end{minipage}
\end{center}
\end{figure*}

\subsection{Adding More Hypercuboids}
We conduct experiments to verify the effectiveness of the proposed hypercuboid variants.  Table \ref{morehypercuboid} shows the performance with varying number of hypercuboids. Notably, adding more hypercuboid can further enhance the model performance, hitting a new state-of-the-art, which reaffirms the importance of modeling the diversity of user interests. Generally, multiple independent hypercuboids usually marginally outperforms multiple concentric hypercuboids. Using four to six hypercuboids can usually lead to a satisfying improvement. We also tried eight and ten hypercuboids but performance degradation is observed.  Note that these two variants will not incur large increases in the number of parameters as increasing the embedding size. We can choose the number of hypercuboids base on the degree of diversity of users' interests in general.

% Please add the following required packages to your document preamble:
% \usepackage{multirow}
\begin{table*}[t]
\caption{Performance (Measure@10) of multiple hypercuboids.}
\begin{tabular}{|c|c|c|c|c|c|c|c|}
\hline
\multicolumn{1}{|l|}{\multirow{2}{*}{}} & \multicolumn{1}{l|}{\multirow{2}{*}{\begin{tabular}[c]{@{}l@{}}Number of\\ Hypercuboids\end{tabular}}} & \multicolumn{3}{l|}{Concentric Hypercuboids} & \multicolumn{3}{l|}{Independent Hypercuboid} \\ \cline{3-8} 
\multicolumn{1}{|l|}{}                  & \multicolumn{1}{l|}{}                                                                             & Recall@10        & NDCG@10          & MAP@10          & Recall@10        & NDCG@10          & MAP@10          \\ \hline
\multirow{5}{*}{E-commerce}   & 2                                                                                                 & 0.1396       & 0.1300        & 0.0802      & 0.1476        & 0.1336        & 0.0830       \\ \cline{2-8} 
 & 3                                                                                                 & 0.1512        & 0.1402        & 0.0876       & 0.1501        & 0.1356        & 0.0846       \\ \cline{2-8} 
                                        & 4                                                                                                 & \textbf{0.1529}        & \textbf{0.1415}        & \textbf{0.0885 }      & 0.1525        & 0.1374        & 0.0859       \\ \cline{2-8} 
                                        & 5                                                                                                 & 0.1508        & 0.1392        & 0.0868       & 0.1529        & 0.1381        & 0.0866       \\ \cline{2-8} 
                                        & 6                                                                                                 & 0.1436        & 0.1334        & 0.0827       & \textbf{0.1530 }       & \textbf{0.1381 }       & \textbf{0.0866 }      \\ \hline
\multirow{5}{*}{Alibaba Dataset V2}             & 2                                                                                                 & 0.0722        & 0.1032        & 0.0480       & 0.0766        & 0.1119        & 0.0534       \\ \cline{2-8} 
   & 3                                                                                                 &  0.0746        &0.1072      & 0.0504       & 0.0774        & 0.1130        & 0.0540       \\ \cline{2-8} 
& 4                                                                                                 &  0.0760       &  0.1095        & 0.0519       & 0.0779        & 0.1136        & 0.0544       \\ \cline{2-8} 
& 5                                                                                                 &  0.0676      & 0.0966       &  0.0453       & 0.0781        & 0.1138        & 0.0546       \\ \cline{2-8} 
& 6                                                                                                 &\textbf{0.0763}        & \textbf{ 0.1100 }       & \textbf{0.0522}      & \textbf{0.0785}        & \textbf{0.1148}        & \textbf{0.0552}       \\ \hline
\end{tabular}
\label{morehypercuboid}
\end{table*}

% Please add the following required packages to your document preamble:
% \usepackage{multirow}
\begin{table}[h]
\caption{Ablation analysis of the neural architectures on three datasets. $\Delta_{\text{Defaut}}$ and $\Delta_{\text{2ndbest}}$ are the relative improvement(\%) over the default version (in Table \ref{allresults}) and the second best baseline, respectively.}
\begin{tabular}{|c|c|c|c|c|}
\hline
                                                                            & Measure@10 & Remove NNs & $\Delta_{\text{Defaut}}$ &$\Delta_{\text{2ndbest}}$  \\ \hline
Books                                                    & Recall    & 0.0558 &  -6.84\%   & +16.25\% \\ \cline{2-5} 
 & NDCG      & 0.0360    &  -6.01\%  & +1.41\% \\ \cline{2-5} 
 & MAP       & 0.0264   &  -8.97\%   & +10.00\% \\ \hline
E-commerce & Recall    & 0.1384  & -0.75\%    & +4.86\% \\ \cline{2-5} 
 & NDCG      & 0.1277   &  -1.00\%   & +10.10\% \\ \cline{2-5} 
   & MAP       & 0.0782   &   -0.84\%  & +11.48\% \\ \hline
\multirow{2}{*}{ \begin{tabular}[c]{@{}c@{}}Alibaba \\  Dataset V1\end{tabular}} & Recall    & 0.0281 &   -6.02\%     &  +3.54\% \\ \cline{2-5} 
 & NDCG      & 0.1174   &  -6.45\% & +2.94\%\\ \cline{2-5} 
 & MAP       &  0.0563  &  -7.70\% & +0.90\% \\ \hline
 \multirow{2}{*}{ \begin{tabular}[c]{@{}c@{}}Alibaba \\  Dataset V2\end{tabular}} & Recall    & 0.0644 & -3.45\%        & -0.77\% \\ \cline{2-5} 
 & NDCG      &  0.0909   &  -4.01\% &-2.26\%\\ \cline{2-5} 
 & MAP       &  0.0414  & -3.94\%& -2.59\% \\ \hline

\end{tabular}
\label{ablation}
\end{table}
\subsection{Using Hypercuboids Only}
We study the impacts of the neural architecture on the model performance. To this end, we report the performance by replacing all the neural architectures with a simple average pooling over the sequences in Table \ref{ablation}, Firstly, we observe that removing the neural architectures will decrease the performance, especially on the Amazon Books dataset and company dataset V1/V2, which confirms the importance of the neural architectures. Next, we note that, even without the neural architectures, the pure hypercuboid architecture still get higher performance than the second best baselines (on Books, E-commerce, and Company Dataset V1), which signifies that the hypercuboid is essential for the improvement. 

\subsection{Sensitivity of Hyper-parameters}
We conduct hyper-parameters analysis on three public datasets.
\subsubsection{Embedding Size} Here, we analyze the impact of the embedding size in Figure \ref{fig:impacthyperparam} (a), (b), and (c). It is clear that increasing the embedding size improves the performance. Unlike models such as Caser, SASRec, and HGN (checking the embedding size analysis in the corresponding papers.),  our model is less likely to be over-fitting with larger dimension size. It suggests that setting a dimension size no less than 100 to be a reasonable choice for our model.

\subsubsection{Sequence length} 
\begin{figure}[t]
\begin{center}
\caption{Impact of the hyper-parameters.}

\begin{minipage}[t]{4.2cm}
\includegraphics[width=4.2cm]{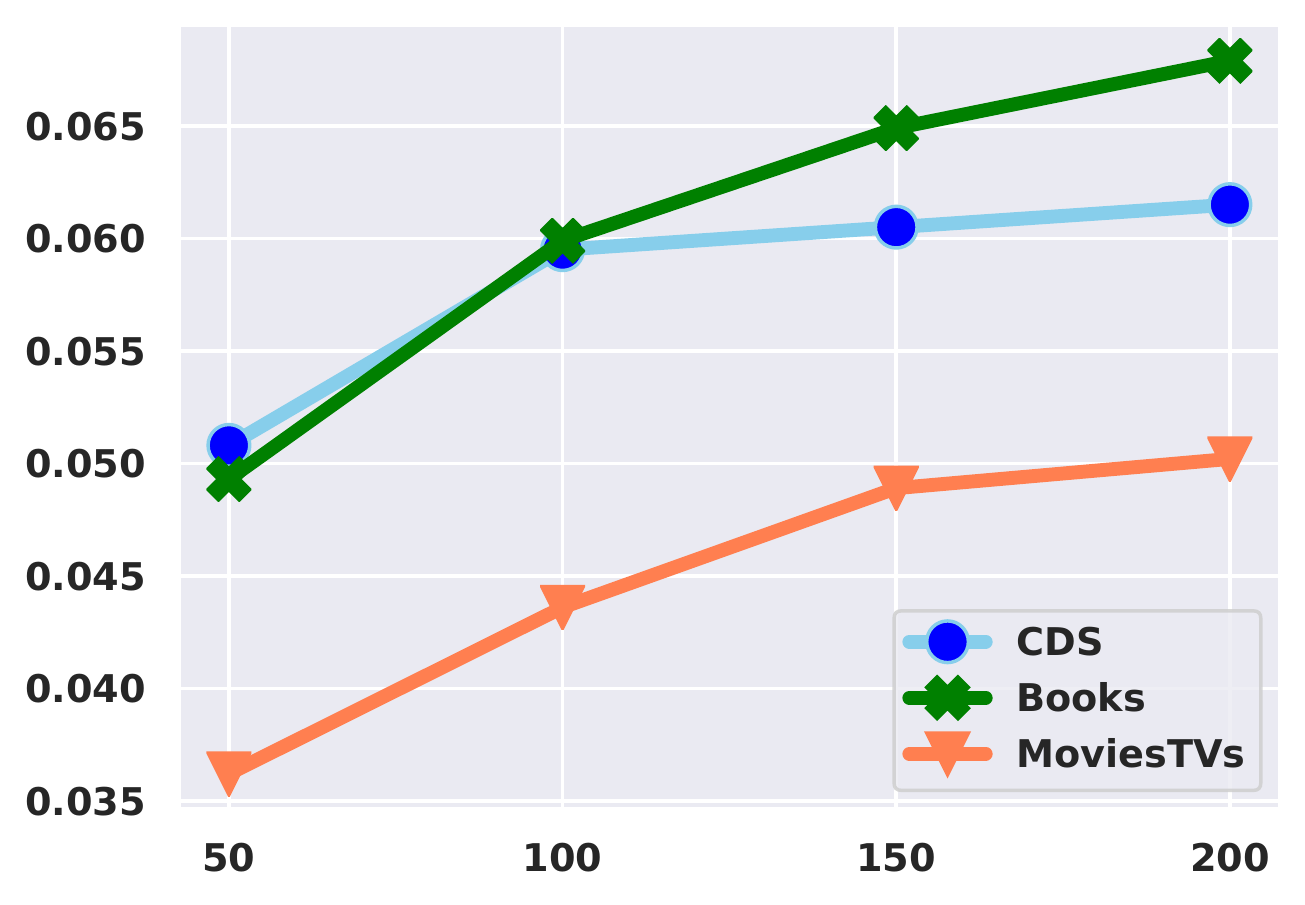}
\centering{(a) Recall@10 with varying $d$ }
\end{minipage}
\begin{minipage}[t]{4.2cm}
\includegraphics[width=4.2cm]{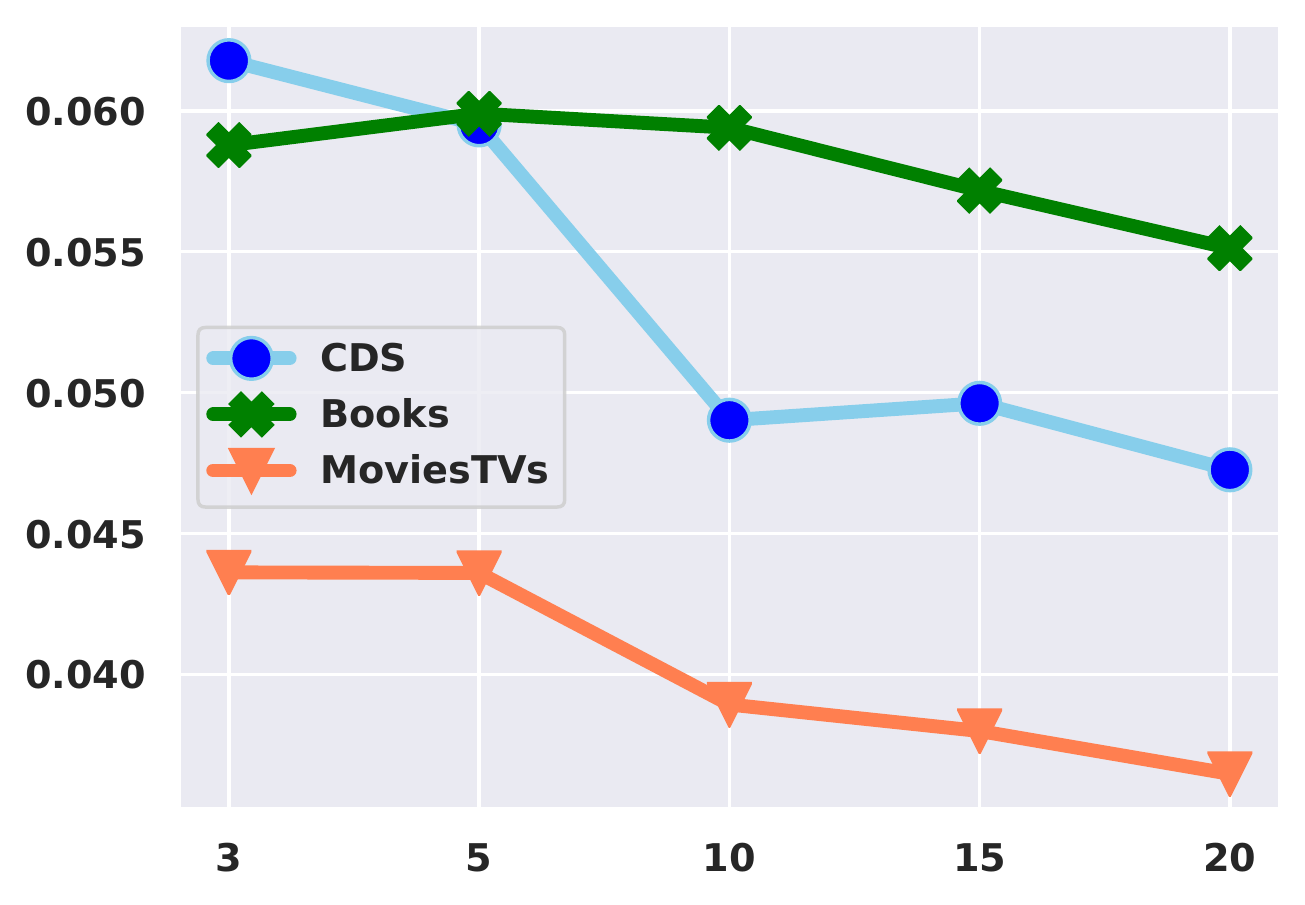}
\centering{(1) Recall@10 with varying $L$  }
\end{minipage}

\begin{minipage}[t]{4.2cm}
\includegraphics[width=4.2cm]{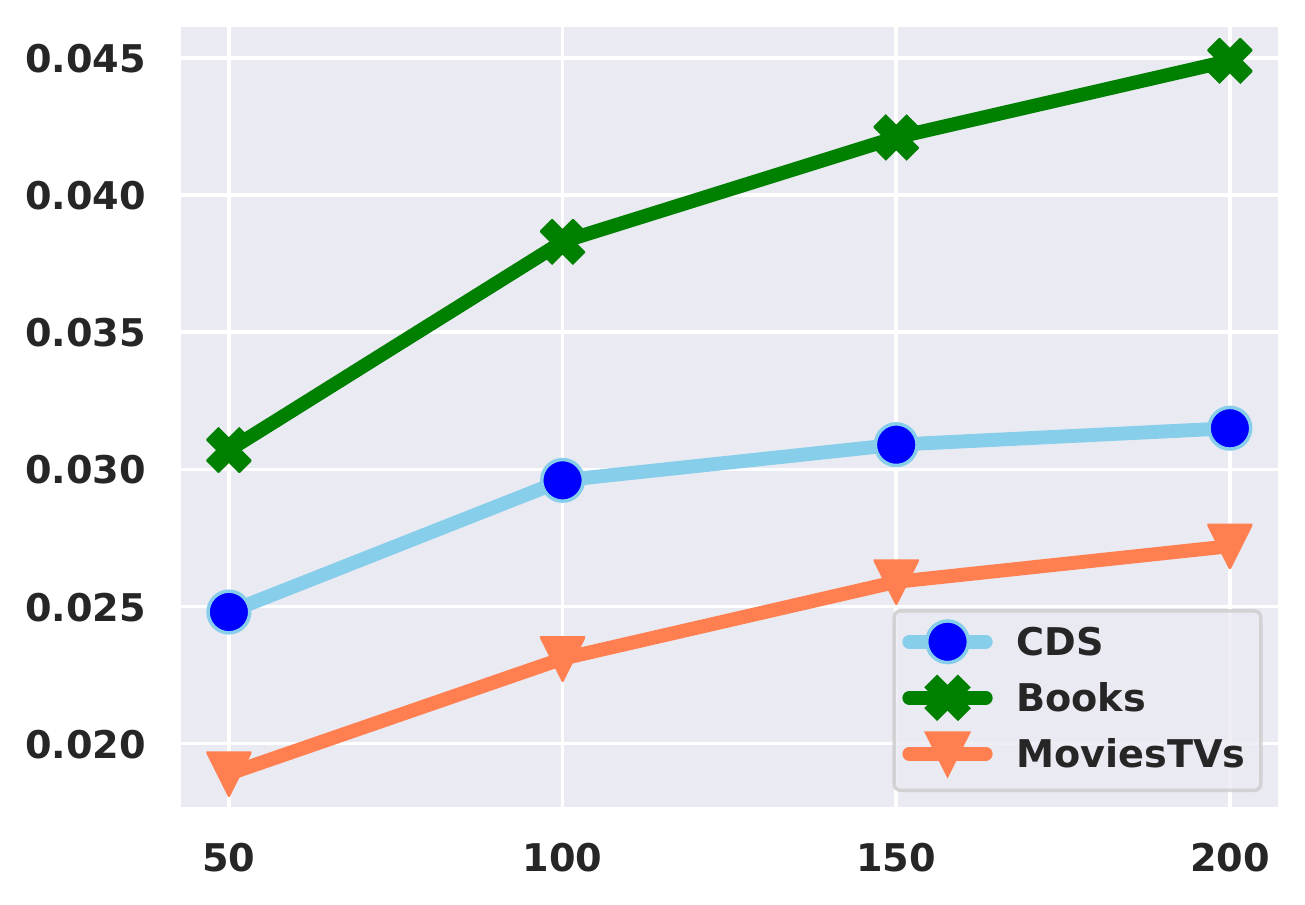}
\centering{(b) NDCG@10 with varying $d$ }
\end{minipage}
\begin{minipage}[t]{4.2cm}
\includegraphics[width=4.2cm]{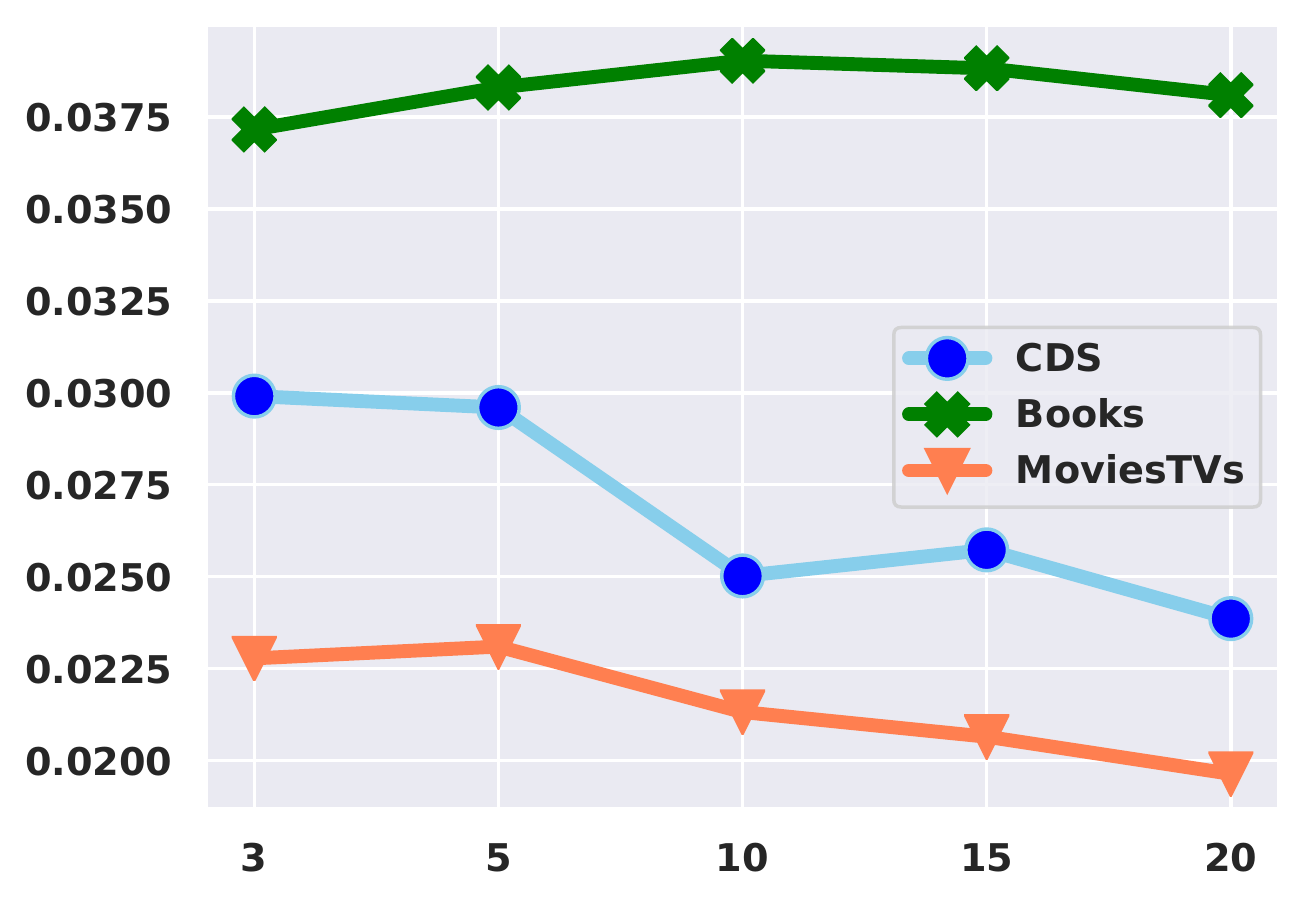}
\centering{(2) NDCG@10 with varying $L$  }
\end{minipage}

\begin{minipage}[t]{4.2cm}
\includegraphics[width=4.2cm]{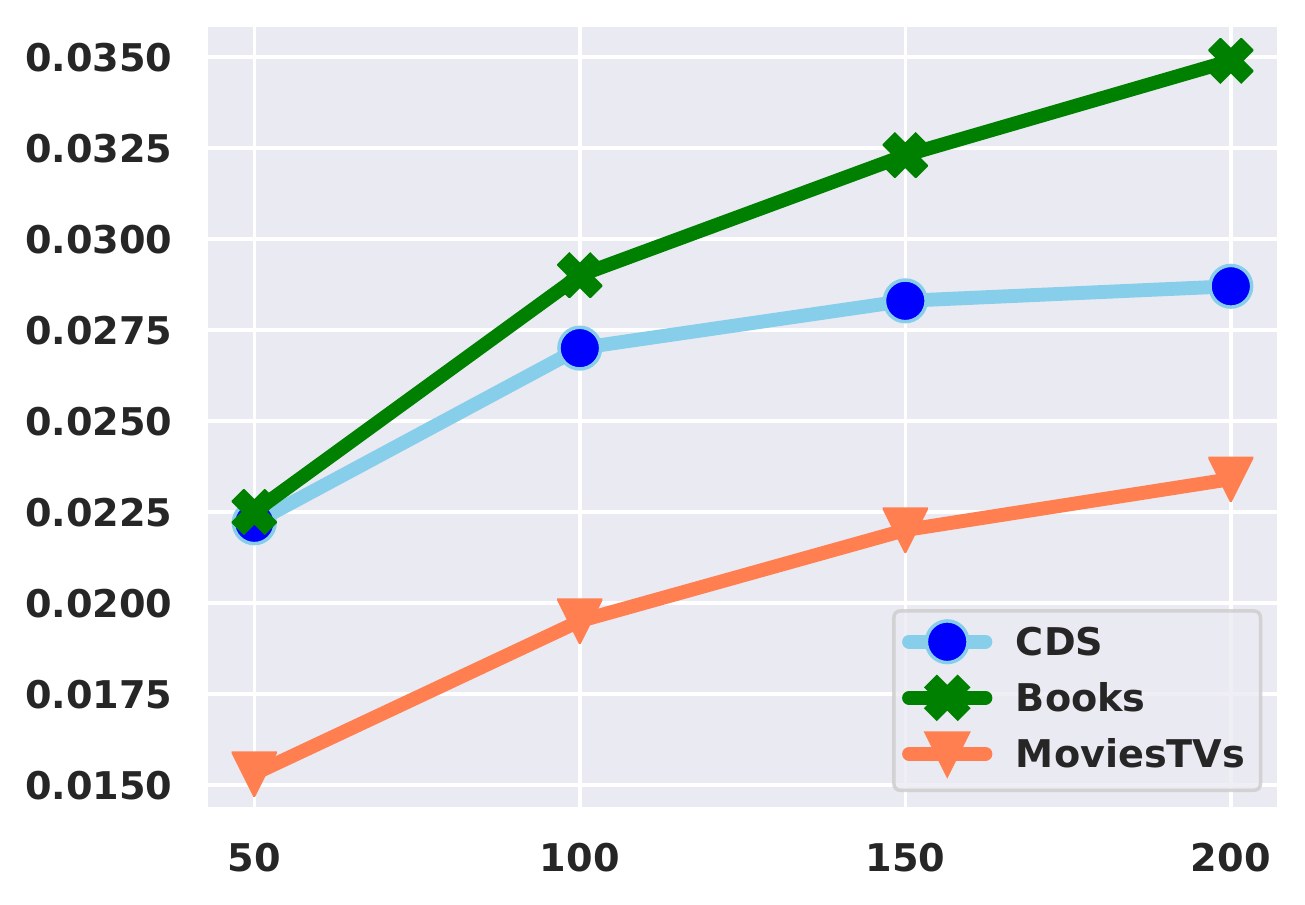}
\centering{(c) MAP@10 with varying $d$ }
\end{minipage}
\begin{minipage}[t]{4.2cm}
\includegraphics[width=4.2cm]{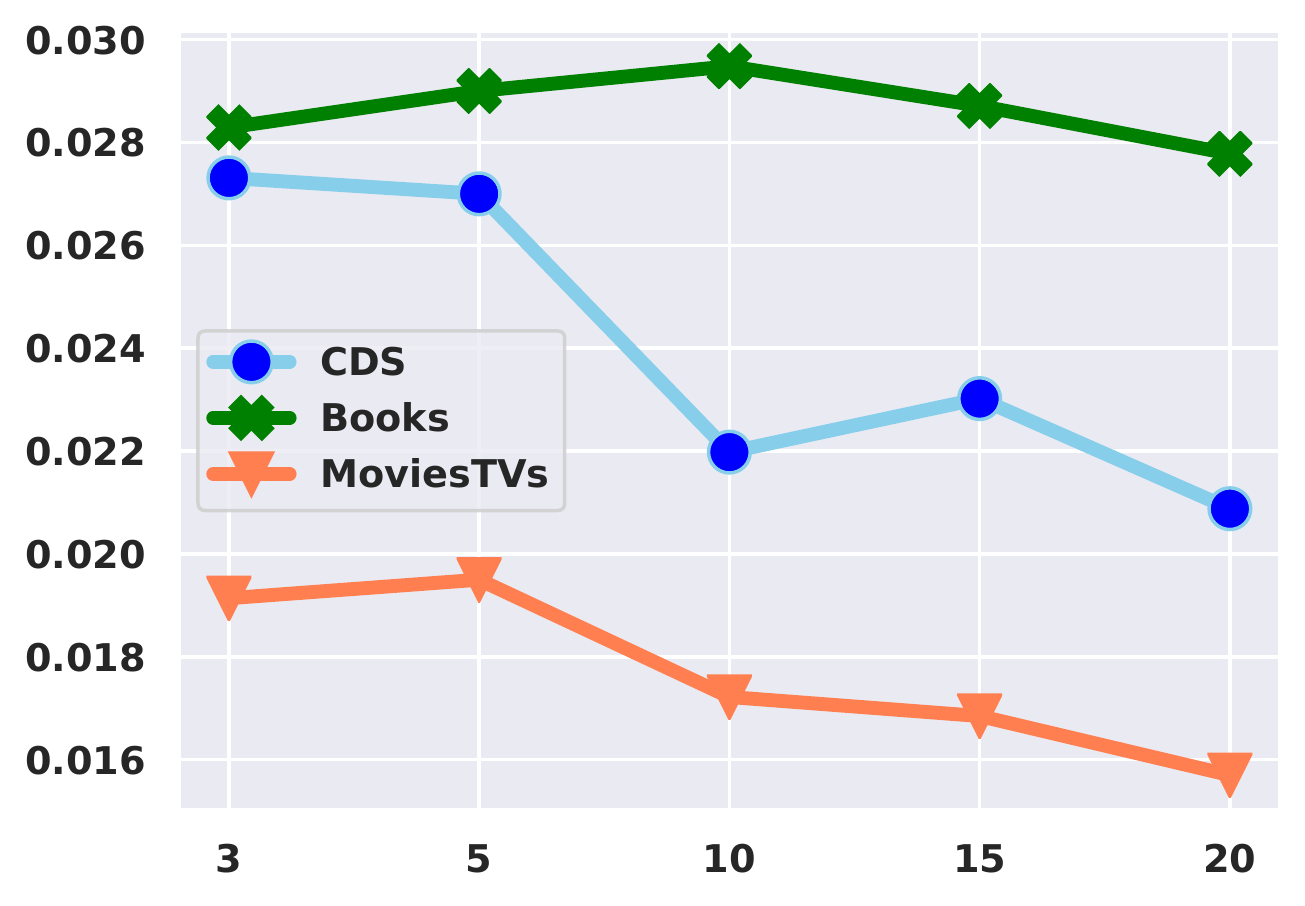}
\centering{(3) MAP@10 with varying $L$ }
\end{minipage}

\label{fig:impacthyperparam}
\end{center}
\end{figure}

By varying the sequence length $L$, we get Figure \ref{fig:impacthyperparam} (1), (2), and (3). We observe that the model gets better performance with shorter sequence length on CDs and Movies\&TVs, while it is more stable on the Books datasets. Overall, it is more preferable to set $L$ to a value around 5 as longer sequence might introduce too many irrelevant activities in the sequences.

\subsubsection{Pooling Function}

Table \ref{tab:pooling} shows the results with different pooling functions. We find that all the three choices (sum, min, and max pooling) lead to deprovements over the mean pooling on the three datasets. In addition, there is no clear winner between the three pooling functions.

% Please add the following required packages to your document preamble:
% \usepackage{multirow}
\begin{table}[h]
\caption{Effect of pooling functions on three datasets}
\begin{tabular}{|c|c|c|c|c|}
\hline
                          & Measure@10 & Sum    & Min    & Max    \\ \hline
CDs     & Recall    & 0.0542 & 0.0529 & 0.0528 \\ \cline{2-5} 
                          & NDCG      & 0.0282 & 0.0268 & 0.0267 \\ \cline{2-5} 
                          & MAP       & 0.0251 & 0.0241 & 0.0235 \\ \hline
Books    & Recall    & 0.0546 & 0.0574 & 0.0570 \\ \cline{2-5} 
                          & NDCG      & 0.0382 & 0.0370 & 0.0370 \\ \cline{2-5} 
                          & MAP       & 0.0272 & 0.0275 & 0.0274 \\ \hline
MovisTVs & Recall    & 0.0369 & 0.0377 & 0.0370 \\ \cline{2-5} 
                          & NDCG      & 0.0212 & 0.0209 & 0.0205 \\ \cline{2-5} 
                          & MAP       & 0.0166 & 0.0165 & 0.0162 \\ \hline
\end{tabular}
\label{tab:pooling}
\end{table}

\begin{figure}[t]
\centering
\includegraphics[width=0.75\linewidth]{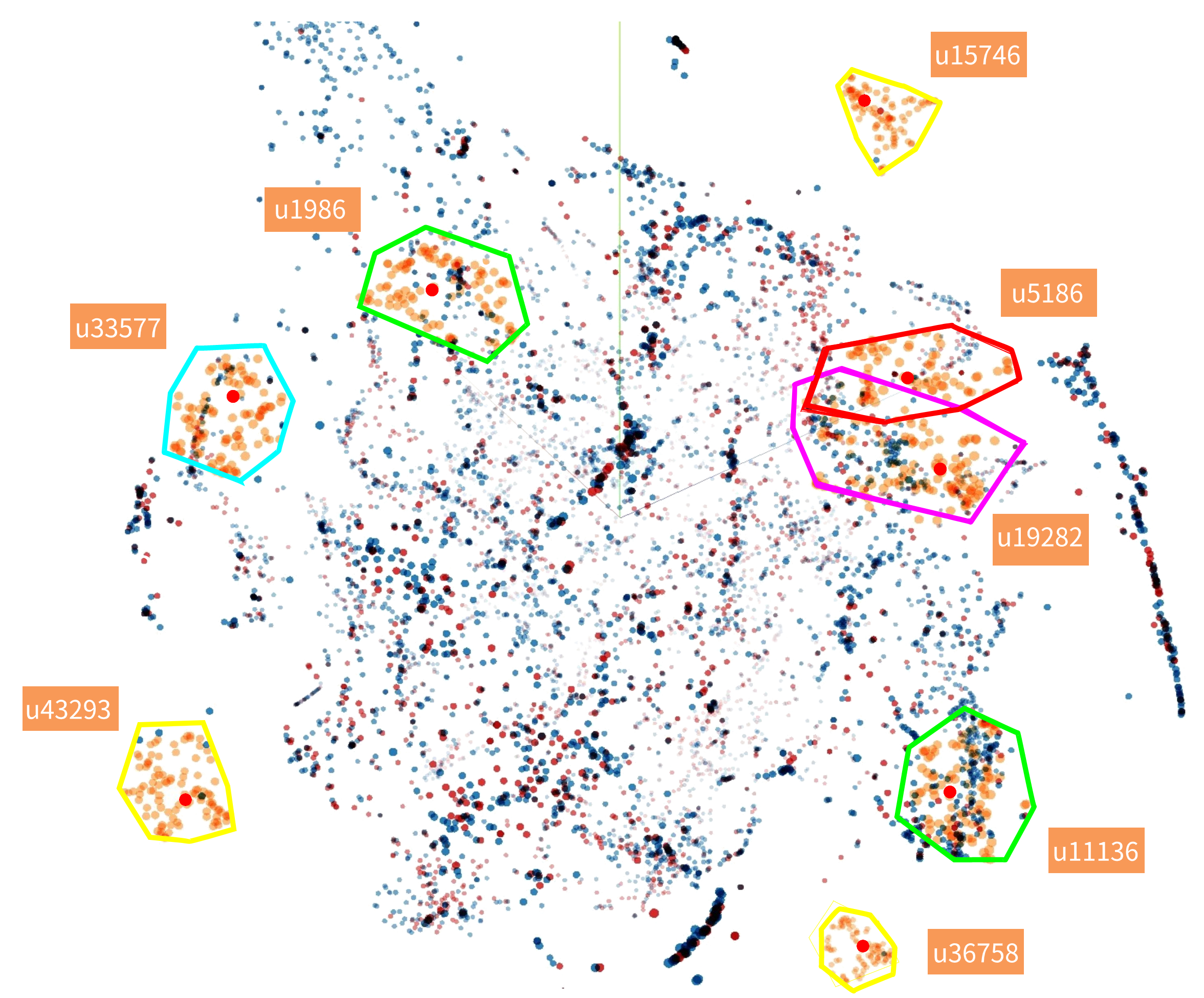}
\centering\caption{Visualization of the user hypercuboids on Books. We randomly show the hypercuboids of eight users. Blue points represent items and red points represent the user centers and the orange points are randomly sampled points in the user hypercuboids. We manually circle each hypercuboid with an irregular polygon.}
\label{fig:visualhypercubioid}
\end{figure}
\subsection{Visualization of User Hypercuboids and Item Embeddings}
In order to check what the proposed model is learning, we visualize the user hypercuboids on the Amazon books dataset. We project 3,000 items and 3,000 user centers into a 3-dimensional space with T-SNE~\cite{maaten2008visualizing} (Perplexity 9, learning rate 10). Then, we randomly choose eight users and, for each of the eight users, we sample 100 points in the hypercuboid uniformly and project them into the same space. Figure \ref{fig:visualhypercubioid} shows the learned hypercuboids. Clearly, our model successfully learned a hypercuboid with different sizes and positions for each user. Items may be located inside or outside the user hypercuboid. Items which are inside the user hypercuboid are more likely to be recommended to this user. Intersections between users' hypercuboids can also be observed.

\begin{figure}[t]
\centering
\includegraphics[width=0.75\linewidth]{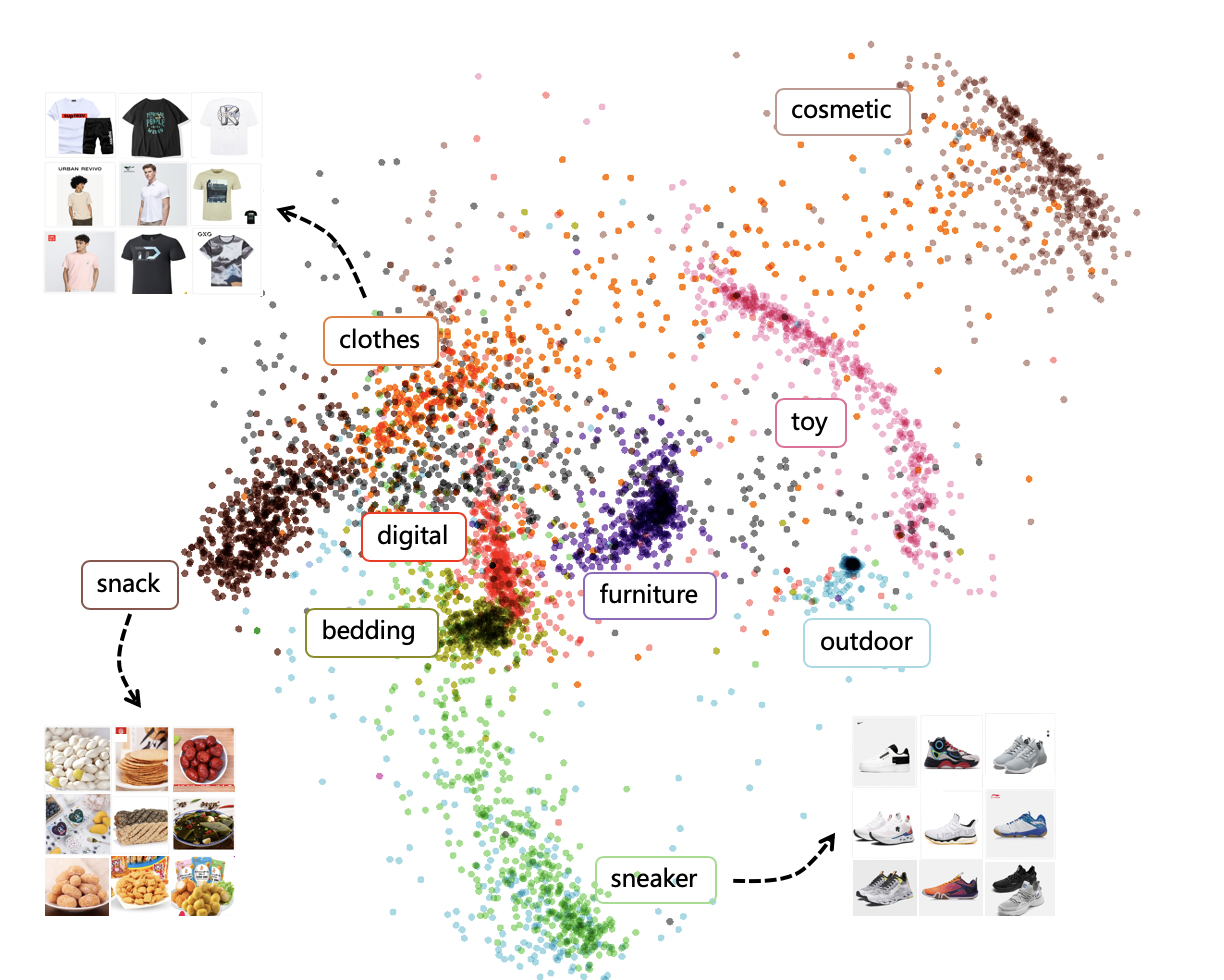}
\centering\caption{Visualization of Item Embeddings on the Alibaba dataset.}
\label{fig:itememb}
\end{figure}

Moreover, our model can also learn effective item representations. To demonstrate this, we project the item embeddings we learned from for the company dataset (V1) into 2-D space. As shown in Figure \ref{fig:itememb},  we observe that items from different categories (e.g., snack, clothes, toy, furniture, etc. ) are roughly separated. Related categories tend to be closer to each other (e.g., bedding and furniture).

\subsection{Case Study with Multiple Hypercuboids}
\begin{figure}[t]
\centering
\includegraphics[width=0.75\linewidth]{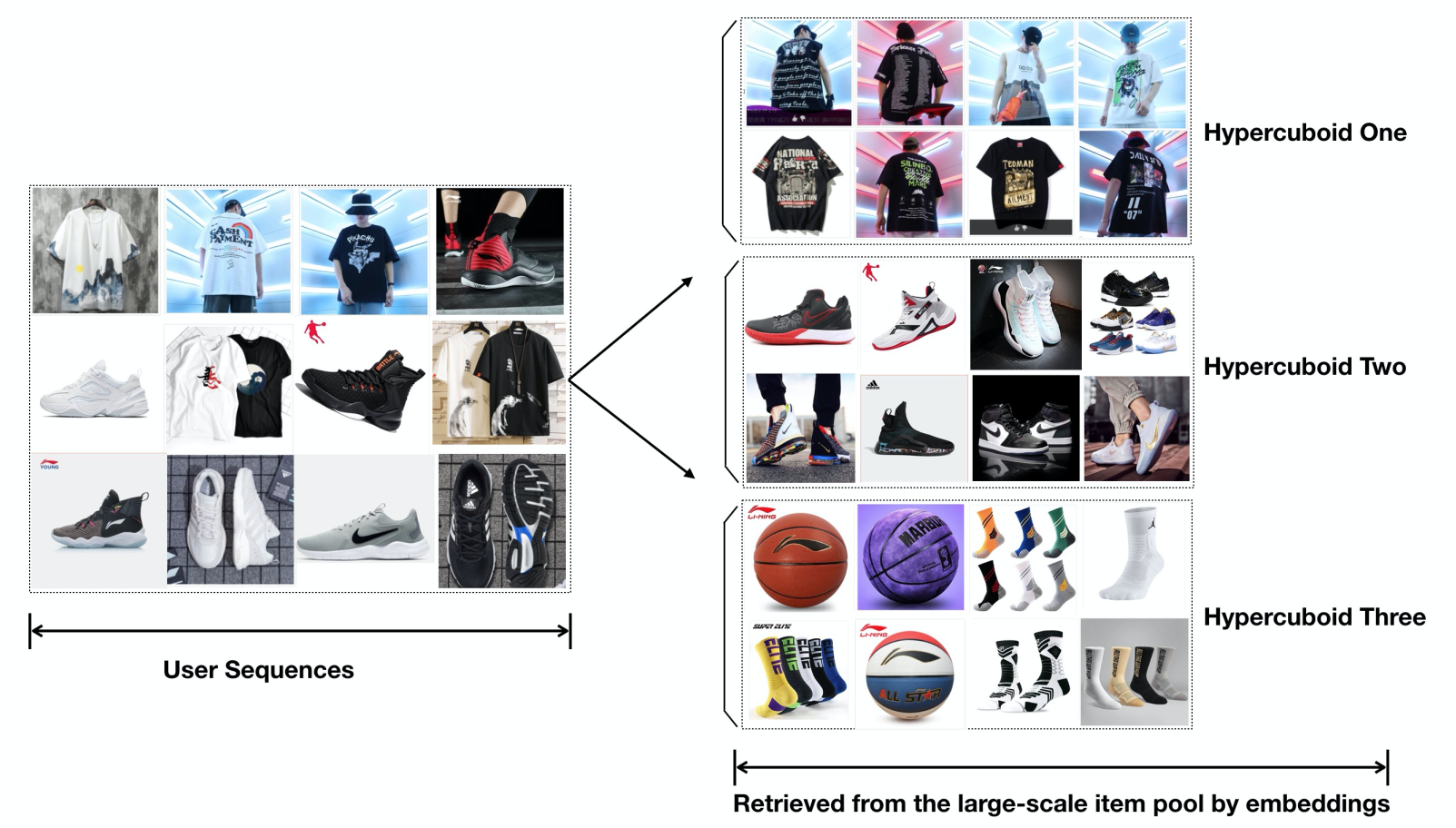}
\centering\caption{Recall results with multiple independent Hypercuboids on the Alibaba dataset.}
\label{fig:case}
\end{figure}

To demonstrate the efficacy of using multiple hypercuboids, we show the recommendations list for a randomly selected user in Figure \ref{fig:case}. In this case, we learn three independent hypercuboids for this user. Several observations can be made: (1). The recalled items are strongly correlated with the corresponding user sequence; (2) Different hypercuboid can capture different aspect of the interests in the historical sequences, i.e, hypercuboid one recalls T-shirts, hypercuboid two recalls sport footwear; (3). A single hypercuboid can also recall a diverse set of items from closely related categories. For example, hypercuboid three recalls both socks and basketballs. Clearly, our model is suitable for capturing user's diverse interests.

% At serving time, items similar to user interests are retrieved by nearest neighbor search. We visualize the
% distribution of these items recalled by each interest based on their
% similarity to the corresponding interest

% To demonstrate the efficacy of the proposed interests interaction method, 
% We count the common items between users, compute the Jaccard similarity and the hypercuboid similarity with the definition \ref{simhypercuboid}. We observe that the hypercuboid similarity scores are consistent with the Jaccard scores in general, which suggests that this might serve as an auxiliary measurement for user similarities. For the user pair $5,186$ and $11,136$, they have zero common items but the Hypercuboid similarity is nonzero. One possible reason is that they share some neighbors with similar interests. We will explore a practical use case of this method in the future.

% \begin{table}[h]
% \caption{Measure the user interests interactions. }
% \begin{tabular}{|c|c|c|c|c|}
% \hline
% User ID & User ID & \begin{tabular}[c]{@{}c@{}}\# Common\\  Items\end{tabular} & Jaccard & \begin{tabular}[c]{@{}c@{}}Hypercuboid \\ Similarity\end{tabular} \\ \hline
% 5186    & 5186    & 286             & 1.0000  &     1.0000                   \\ \hline
% 5186    & 19282   & 79              & 0.1904  &   0.4653                     \\ \hline
% 5186    & 1986    & 47              & 0.1260  &  0.4561                      \\ \hline
% 5186    & 11136   & 0               & 0.0000  &    0.3349                    \\ \hline
% 5186    & 33577   & 40              & 0.1028  &      0.2260                  \\ \hline
% 5186    & 43293   & 0               & 0.0000  &    0.0979                    \\ \hline
% \end{tabular}
% \label{tab:hypersim}
% \end{table}

\section{Related Work}

Recommender systems have been widely studied in the literature~\cite{ricci2011introduction,koren2008factorization,koren2009matrix,adomavicius2005toward,10.1145/2783258.2783273,shi2014collaborative,su2009survey}. Conventional recommendation methods usually represent users and items as points/vectors in a latent space. The interactions are modelled with vectors similarity~\cite{koren2009matrix,hu2008collaborative,zhang2019quaternion} or points closeness~\cite{hsieh2017collaborative,vinh2020hyperml}. In the recent years, we can easily observe the increasing numbers of neural network models for recommendation~\cite{zhang2019deep,zhang2019dive}. A number of works~\cite{he2017neural,covington2016deep,wu2016collaborative,xue2017deep} proposed to model the interactions with multilayered perceptrons. There is also an emerging line of works focusing on capturing the sequence patterns in user activities to enrich user representations~\cite{tan2016improved,feng2015personalized,quadrana2018sequence}. Hidasi et al.~\cite{hidasi2015session,hidasi2016parallel} proposed a recurrent neural network based model for session-based recommendation. Convolutional neural networks based sequence modeling approaches show promising results on sequential recommendation~\cite{tang2018personalized,yuan2019simple}. Self-attention mechanism~\cite{kang2018self,zhou2018atrank,huang2018csan} can also utilized to enhance the performance by emphasizing important activities in user behaviors. The recent model, hierarchical gating networks~\cite{ma2019hierarchical}, moving beyond RNN, CNN, and self-attention, is a gating network that consists of a feature gating module, an instance gating module and an item-item product module. Different from previous works, our model takes the advantages of hypercuboid representations to overcome the limitations of traditional user representation methods. A customized sequence modeling block is also designed to boost the effectiveness of the hypercuboid by incorporating information from user sequential behaviors.

Hypercuboid is an extension of cuboid in higher dimensions. It has attracted attention in the field of machine learning recently~\cite{4912198,maji2012rough,ren2020querybox,zhang20114}. Maji et al.~\cite{maji2012rough} proposed a rough hypercuboid method for relevant and significant features selection. Vilnis et al.~\cite{vilnis-etal-2018-probabilistic} proposed a probabilistic embedding method with hypercuboid lattice Measures for knowledge graphs. In the followed-up work, Li et al.~\cite{li2018smoothing} improved this model to enhance the robustness in the disjoint case while keeping the desirable properties. Ren et al.~\cite{ren2020querybox} proposed a hypercuboid representation approaches for reasoning over knowledge graphs. Our work is also related to Venn Diagrams~\cite{venn1880diagrammatic} in the sense that we can treat users as a set of entities they preferred.

\section{Conclusion}
In this paper, we propose a novel recommendation method with hypercuboids. In our model, users are represented as hypercuboids and the relationship between users and items are modeled with the distance between points and hypercuboids.  It demonstrates the state-of-the-art performance on four publicly available and two commercial datasets for personalized recommendation. To capture the diversity of user interests, we proposed two variants of hypercuboids which could further improve recalling performance. We also studied the impacts of effects of hyperparameters and provided case analysis on real-world dataset.

%% The next two lines define the bibliography style to be used, and
%% the bibliography file.
\bibliographystyle{ACM-Reference-Format}
\bibliography{sample-base}

\end{document}